\documentclass[aps,pre,preprint,numerical,a4paper]{revtex4-1}

\usepackage{graphicx}
\usepackage{calc}
\usepackage{amsmath}
\usepackage{bbm}
\usepackage{amsfonts}
\usepackage{amssymb}
\usepackage{amsbsy}
\usepackage[T1]{fontenc}
\usepackage{theorem}
\usepackage{stmaryrd}
\usepackage{tipa}
\usepackage{textcomp}
\usepackage{subfigure}
\usepackage{epsfig}
\usepackage{lmodern}
\usepackage{framed}
\usepackage[subfigure]{tocloft}
\usepackage{subeqnarray}
\usepackage{alphalph}
\usepackage[refpage,cfg,prefix]{nomencl}
\usepackage{multirow}
\usepackage{xifthen}
\usepackage{enumerate}
\usepackage{color}
\usepackage{wasysym}
\usepackage{bibentry}
\nobibliography*
\usepackage{verbatim}
\usepackage{url}
\usepackage{setspace}
\usepackage{xr}
\usepackage{color,colortbl}
\usepackage{relsize}
\usepackage[toc,page]{appendix}

\makenomenclature

\newcommand{\D}[2]{\frac{\partial #1}{\partial #2}}
\newcommand{\DD}[2]{\frac{\partial^2 #1}{\partial #2^2}}

\newcommand{\listofalgorithms}{\textbf{\Huge{List of Algorithms}}}

\newlistof{Algorithm}{exp}{\listofalgorithms}

\newif\ifletter

\definecolor{light-gray}{gray}{0.4}
\usepackage{epstopdf}
\def\newblock{}
\allowdisplaybreaks
\begin{document}

\title[Incorporating cell-cell pulling into models of cell migration]{Pulling in models of cell migration}
\author{George Chappelle}
\affiliation{Department of Mathematics, Imperial College London, SW7 2AZ, UK}	
\author{Christian A. Yates}\email[Corresponding Author: ]{c.yates@bath.ac.uk}
\homepage{Website: www.kityates.com}
\affiliation{Centre for Mathematical Biology, Department of Mathematical Sciences, University of Bath, Claverton Down, Bath, BA2 7AY, UK}	

\date{\today}

\begin{abstract}
There are numerous biological scenarios in which populations of cells migrate in crowded environments. Typical examples include wound healing, cancer growth and embryo development. In these crowded environments cells are able to interact with each other in a variety of ways. These include excluded volume interactions, adhesion, repulsion, cell signalling, pushing and pulling. 

One popular way to understand the behaviour of a group of interacting cells is through an agent-based mathematical model. A typical aim of modellers using such representations is to elucidate how the microscopic interactions at the cell-level impact on the macroscopic behaviour of the population. At the very least, such models typically incorporate volume exclusion. The more complex cell-cell interactions listed above have also been incorporated into such models; all apart from cell-cell pulling.

In this paper we consider this under-represented cell-cell interaction, in which an active cell is able to `pull' a nearby neighbour as it moves. We incorporate a variety of potential cell-cell pulling mechanisms into on- and off-lattice agent-based volume exclusion models of cell movement. For each of these agent-based models we derive a continuum partial differential equation which describes the evolution of the cells at a population-level. We study the agreement between the agent-based models and the continuum, population-based models, and compare and contrast a range of agent-based models (accounting for the different pulling mechanisms) with each other. We find generally good agreement between the agent-based models and the corresponding continuum models that worsens as the agent-based models become more complex. Interestingly, we observe that the partial differential equations that we derive differ significantly, depending on whether they were derived from on- or off-lattice agent-based models of pulling. This hints that it is important to employ the appropriate agent-based model when representing pulling cell-cell interactions. 
\end{abstract}

\maketitle

\section{Introduction}

Cell migration is an important process in the development and maintenance of multi-cellular organisms. The physical mechanisms by which individual cells propel themselves have been well studied, \citep{lauffenburger1996cmp,ridley2003cmi}. The most common such mechanism amongst eukaryotic cells, known as amoeboid movement, is characterised by frequent changes in cell shape resulting in a crawling motion \citep{van2011acu,condeelis1977cba}.

Recently, interest has turned to understanding the migration of large populations of cells in crowded environments. Such crowded migration is observed in a variety of biological processes, including in embryo development \citep{kulesa1998ncc,keller2005cmd}, wound healing \citep{poujade2007cme,deng2006rma} and cancer growth \citep{friedl2009ccm}. In such crowded environments cell-cell interactions are inevitable. These interactions come in many forms, from physical exclusion forces \citep{simpson2010mbc}, to chemical signalling \citep{weijer2004dm}, to pushing \citep{chen2013mcc,schmidt2010icm}, to adhesion \citep{anguige2009odm,glazier1993sda,armstrong2009aac}, and even indirect influence through deformations of the extra-cellular matrix \citep{stetler1993tci}. It is also possible for cells that are not in direct contact to interact over large distances by extending long protrusions called filopodia \citep{kulesa1998ncc,kasemeier2005inc}. 

There is an extensive literature on the subject of modelling cell migration in a wide variety of biological contexts. Most of the models proposed fall into two categories. The first type are continuous, deterministic, partial differential equation (PDE) models which describe the average density of cells at a given location and time \citep{painter2002vfq,hillen2008ugp,keller1971tbc,keller1971mfc}. The second type of models are discrete, stochastic, agent-based models (ABMs) which explicitly represent each cell as an autonomous agent. The PDE models are usually faster to simulate and are often amenable to mathematical analysis, however they cannot account for the stochastic and agent-based phenomena inherent to some systems. ABMs, on the other hand, are capable of capturing stochastic effects but are more computationally expensive and must be run multiple times to obtain statistics that characterise the process they are modelling. ABMs are also typically harder to analyse mathematically making simulation a necessity. 

One of the main challenges in characterising these processes is to understand how the microscopic, agent-based interactions manifest themselves in the macroscopic behaviour of the population of cells. It is often possible to derive a PDE model describing the mean-field behaviour of an ABM by taking the appropriate limits in time and space. Specifically, for the majority of the cell-cell interactions mentioned above, multi-scale mathematical modelling and analysis has been performed in order to link ABMs to population-level models and consequently to provide a better understanding of the attendant biological processes  \cite{deroulers2009mtc,fernando2010nde,anguige2009odm,thompson2012mcm,yates2015ipe, penington2011bmm,simpson2010cip,simpson2009mss,treloar2011vjm}. \citet{deroulers2009mtc} present ABMs describing a cell adhesion/repulsion mechanism motivated by observations of the behaviour of tumour cells. \citet{fernando2010nde} generalises this model to encompass many forms of contact forming, maintaining and breaking mechanisms. \citet{penington2011bmm} generalise the interactions even further and to arbitrary dimensions. \citet{yates2015ipe} consider an ABM in which cells are able to push other cells out the way. In each of these papers non-linear diffusion equations are derived as the corresponding continuum models. By analysing the resulting PDEs it is possible to better understand population-level characteristics of the original ABMs. However, it should be noted that the mean-field continuum models are generally  an approximation to the population dynamics of the underlying ABM and, as such, should be treated with caution.

The studies cited above all begin with on-lattice ABMs. Unfortunately, there is evidence that the lattice structure may produce unintended artefacts in the population-level behaviour \cite{flache2001dig, wolfram1986caf}, including, most noticeably, anisotropy. Some recent studies have therefore considered off-lattice ABMs in which cells are able to move on a continuous domain \cite{dyson2012mli, dyson2014ive, yates2015ipe}. These off-lattice models are typically more computationally intensive to simulate and the range of scenarios for which corresponding mean-field continuum models can be derived is also generally more limited. 

An important cell-cell interaction which has been overlooked in the modelling literature is one which we refer to as cell-cell pulling. This phenomena is characterised by an active cell pulling, dragging (or otherwise causing to follow) another cell behind it as it moves. Some form of cell-cell pulling has been observed in many of the prototypical examples of populations undergoing collective cell migration. \citet{yamanaka2014iva}, for example, studied the behaviour of zebrafish pigment cells \textit{in vitro}. A `chase-run' interaction was observed between two types of migrating cells (xanthophores and melanophores). The interaction between the two cells only happens when they are very close to each-other. Another example of cell-cell pulling is in the collective migration of neural crest cells. \citet{mclennan2012mmc} studied the long range migration of embryonic chick neural crest cells. They showed that some cells act as leaders (or trailblazers) causing other cells to follow them. In contrast to the previous example, the pulling mechanism considered in this study can act over larger distances mediated by cellular extensions (filopodia). This study also demonstrated that a single leader cell can pull multiple follower cells. 

Similar phenomena have been observed in other biological contexts, although these are often referred to by different names. For example during \textit{Drosophila} oogenesis \citep{bianco2007tdm}, mouse mammary gland development \citep{ewald2008cem}, zebrafish lateral line development \citep{ghysen2007llm}, wound healing \citep{poujade2007cme,gov2007ccm} and cancer growth \citep{khalil2010dlc}. The exact nature of the pulling mechanism varies from one application to another. For this reason we propose a range of possible models. 


In order to address this relatively neglected area of mathematical modelling we devise and study both on- and off-lattice ABMs of cell-cell pulling. For a range of proposed pulling mechanisms, by taking the diffusive limit of the average occupancy equation, we derive corresponding macroscopic partial differential equations for the population-level behaviour of the cells. We compare the behaviour of these population-level models to that of the ABMs from which they originate  and suggest reasons for their agreement or discrepancy. We find that the population-level models derived from either on- and off-lattice ABMs display significantly different characteristics to each other, consistent with findings from previous studies on alternative cell-cell interactions \citep{yates2015ipe}.

In this paper we begin, in Section \ref{chapter: On lattice Pulling}, by outlining a simple on-lattice ABM of cell-cell pulling and deriving the corresponding mean-field PDE. We compare this basic pulling model with a simple random walk model in order to distinguish the effects of cell-cell pulling. In Section \ref{section:extensions_to_the_pulling_paradigm} we present some extensions to the original pulling model. In particular, we allow multiple agents to be pulled simultaneously, combine models for pushing and pulling and allow cell-cell pulling at a distance. In each case, we derive the corresponding mean-field PDE and compare the average behaviour of the ABMs that of the PDE. In Section \ref{chapter:Off lattice models} we formulate some off-lattice models of cell migration which incorporate cell-cell pulling. We derive continuum PDEs from these models and compare their behaviour to the averaged behaviour of the corresponding ABMs. Finally, in Section \ref{chapter:discussion} we finish with a discussion of the merits of each of the models we have formulated and by placing the work in context, suggest areas to which this work could be extended.

\section{Simple cell-cell pulling} \label{chapter: On lattice Pulling}
In this section we begin by presenting the most basic, on-lattice, stochastic, volume excluding ABM for cell migration in subsection \ref{subsection:individua_based_model}. We then increase the model's complexity by incorporating a pulling mechanism. We also derive the corresponding deterministic, population-level descriptions in subsection \ref{sec:Cont_model} and compare the two models in subsection \ref{subsec:Comparison_with_pushing}.

\subsection{On lattice agent-based model}\label{subsection:individua_based_model}

We initially model cell migration using an on-lattice two-dimensional exclusion process. We refer to the cells as `agents', each of which occupy a single lattice site. A site can be occupied by at most one agent. The lattice sites are square, with length $\Delta$. There are $L_x$ sites along the horizontal and $L_y$ sites in the vertical direction. 
The occupancy of site $(i,j)$ is unity if the site is occupied and zero otherwise.

\begin{figure}[htbp]
\begin{center} 
\subfigure[] {
\includegraphics[height=0.15\textheight]{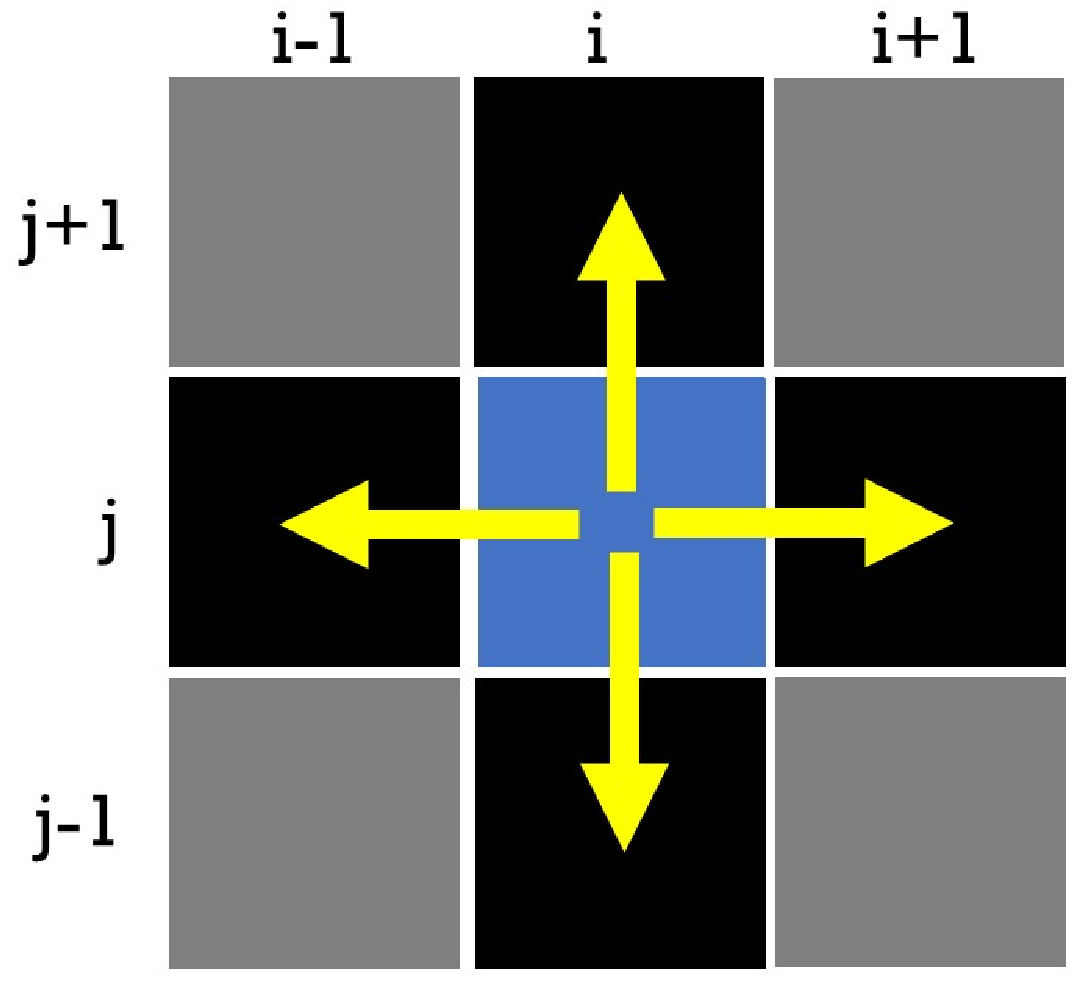}
\label{fig:basic_ex_schematic}
}
\subfigure[]{
\includegraphics[height=0.15\textheight]{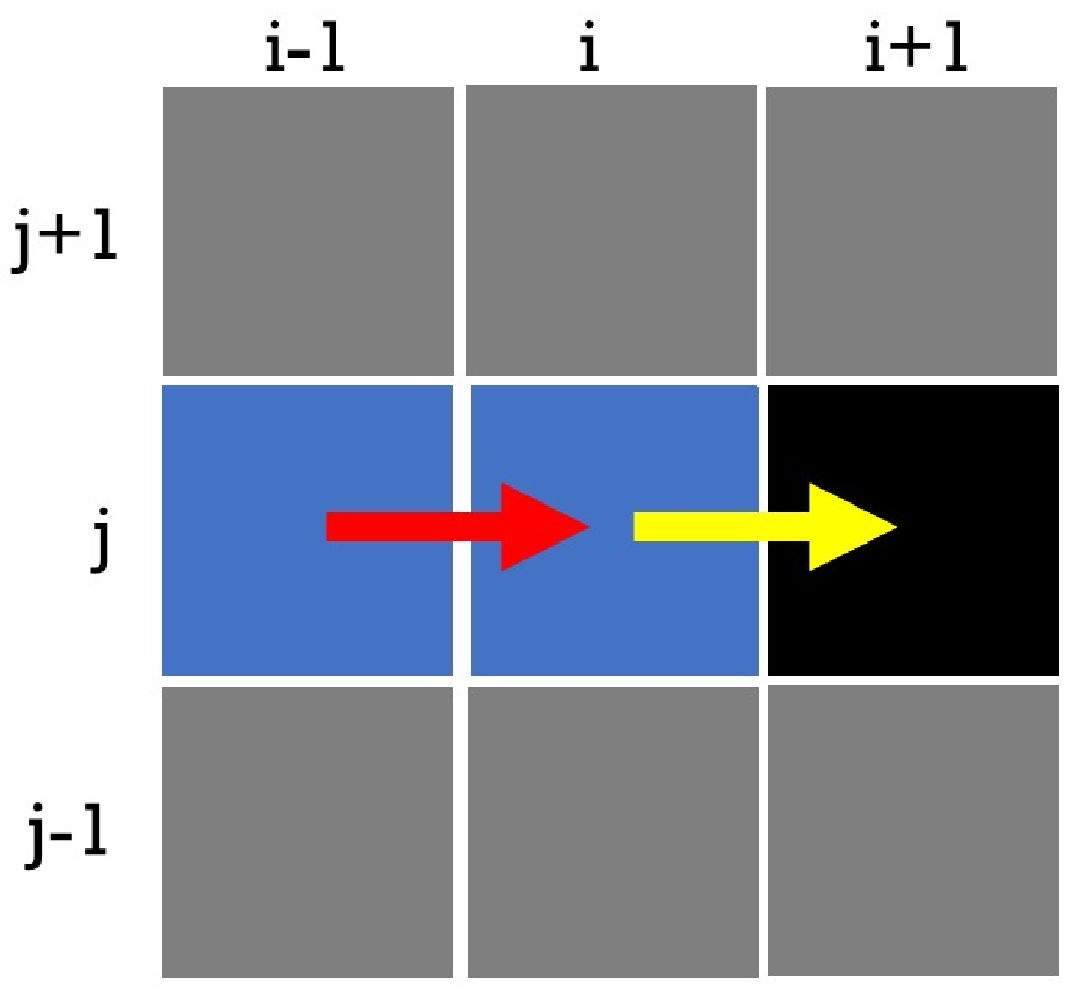}
\label{fig:pulling_schematic}
}
\end{center}
\caption{Occupied sites are blue and vacant site are black. The occupancy of grey sites has no effect on the movements depicted. An active move is denoted by a yellow arrow and a pulling move is denoted by a red arrow. Panel \subref{fig:basic_ex_schematic} shows an agent in site $(i,j)$ which has been selected to move. In this case all four of its neighbours are vacant so it could move to each of them with probability $1/4$. Panel \subref{fig:pulling_schematic} shows cell pulling event. In this case an agent in site $(i,j)$ has been chosen to move rightwards to site $(i+1,j)$ (yellow arrow) and pulls, with probability $w$, an agent in site $(i-1,j)$ into site $(i,j)$ (red arrow).}
\label{fig:cell_schematics}
\end{figure}

We initialise $N$ agents on the lattice, the process then evolves in continuous time as follows. We define a parameter $p$ so that the probability that an agent, chosen uniformly at random, attempts to move sites in the time interval $[t,t+dt]$ is given by $pdt$. Once an agent is selected to move it attempts to jump to any of its four neighbouring sites (see Figure \ref{fig:basic_ex_schematic}) with equal probability. If the chosen site is occupied then the move is aborted. This model is commonly referred to as the simple exclusion process. We use periodic boundary conditions along the horizontal boundaries and reflecting boundary conditions along the vertical boundaries. In this paper we initialise agents in the centre of the domain and set $L_x$ large enough so that the boundary effects at $x=0$ and $x=L_x$ can be ignored. We are therefore effectively modelling cell migration on the surface of a long thin cylinder. 

The simplest case of cell-cell pulling (see Figure \ref{fig:pulling_schematic}) works by adapting the simple exclusion process model as follows. Suppose an agent moves rightwards from site $(i,j)$ to site $(i+1,j)$. If there is an agent in site $(i-1,j)$ then this agent is pulled into site $(i,j)$ with probability $w$. Otherwise, if there is no agent in site $(i-1,j)$ then movement proceeds as in the simple exclusion process. An analogous mechanism applies to agents moving leftwards, upwards and downwards. 

\begin{figure}[htbp]
\begin{center}
\begin{minipage}{.49\textwidth}
\subfigure[]{
\includegraphics[width=1\textwidth]{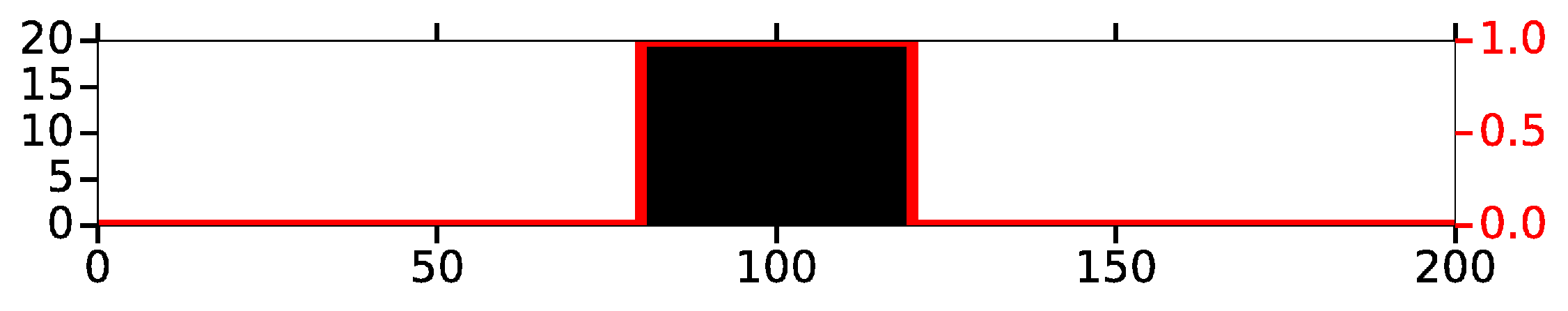}
\label{subfigure:av_no_pulling0}
}
\subfigure[]{
\includegraphics[width=1\textwidth]{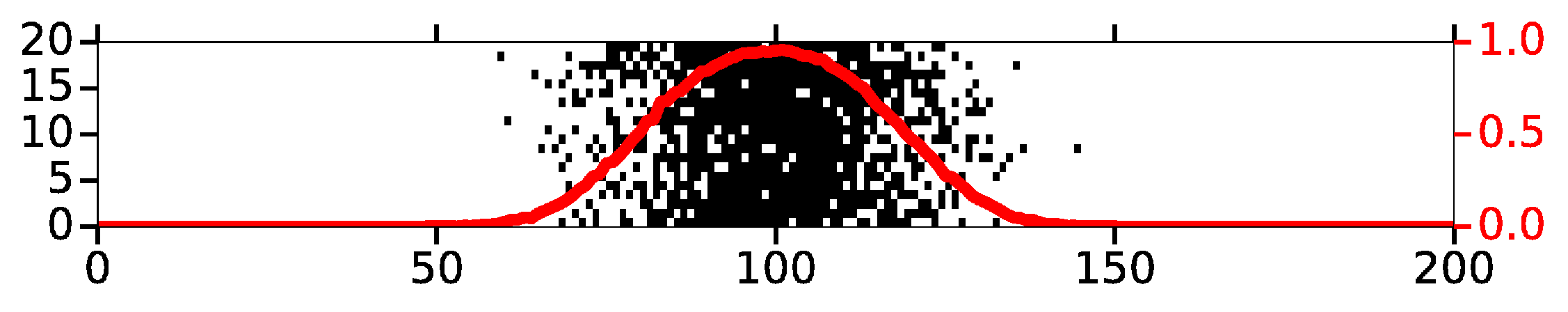}
\label{subfigure:av_no_pulling200}
}
\subfigure[]{
\includegraphics[width=1\textwidth]{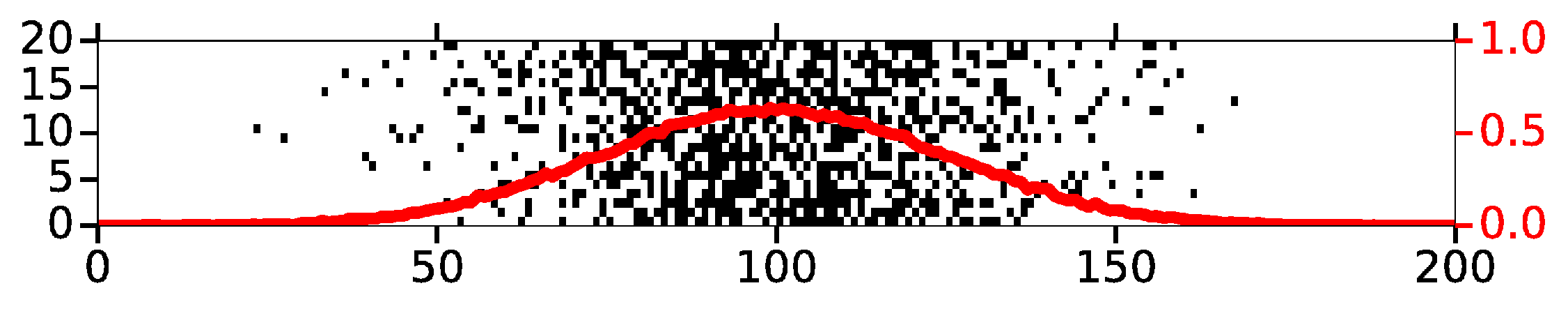}
\label{subfigure:av_no_pulling1000}
}
\end{minipage}
\begin{minipage}{.49\textwidth}
\subfigure[]{
\includegraphics[width=1\textwidth]{./snapshot_av_ex_0_py.eps}
\label{subfigure:av_pulling0}
}
\subfigure[]{
\includegraphics[width=1\textwidth]{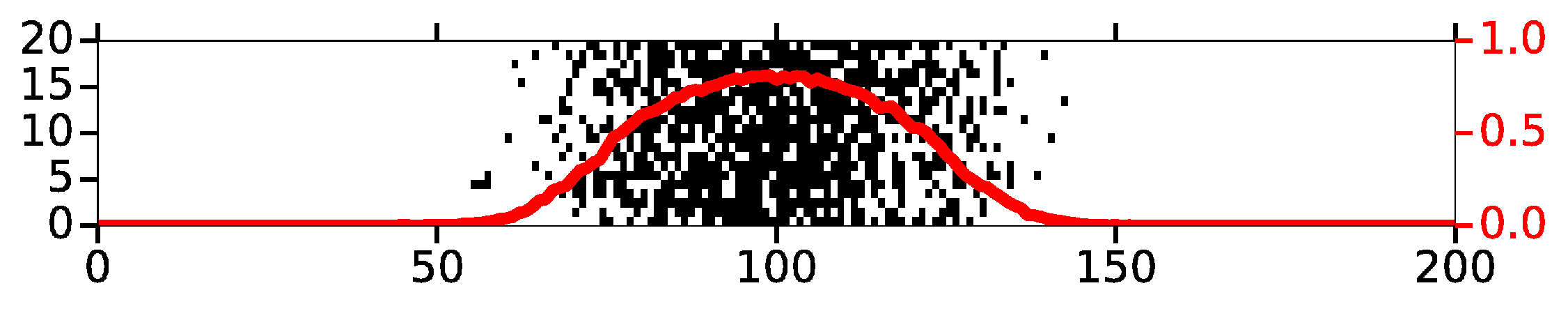}
\label{subfigure:av_pulling200}
}
\subfigure[]{
\includegraphics[width=1\textwidth]{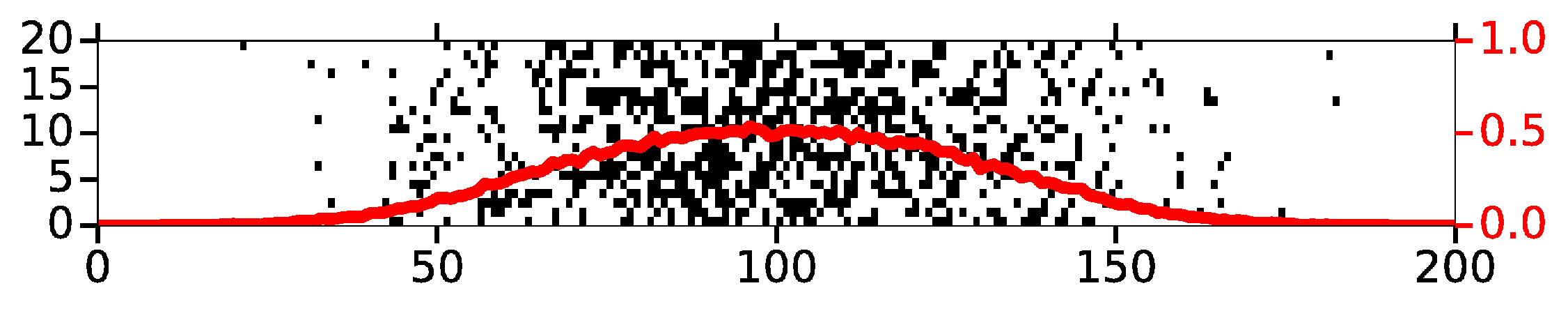}
\label{subfigure:av_pulling1000}
}
\end{minipage}
\end{center}
\caption{The evolution of the lattice occupancy for \subref{subfigure:av_no_pulling0}-\subref{subfigure:av_no_pulling1000} the simple exclusion process model and \subref{subfigure:av_pulling0}-\subref{subfigure:av_pulling1000} the exclusion process model with agent-agent pulling ($w=1$). All simulations are carried out on a lattice with dimensions $L_x=200$, $L_y=20$ and with periodic boundary conditions along the horizontal boundaries. Lattice occupancy is shown at $t=0$, $t=200$ and $t=1000$. The red curve depicts averaged column density over 100 repeats. We initialize the lattice with agents agents occupying all of the sites ($x$,$y$) such that $81 \leq x \leq 120$ and $1 \leq y \leq 20$.}
\label{fig:comparison_pulling_vs_non_pulling}
\end{figure}

In Figure \ref{fig:comparison_pulling_vs_non_pulling} we present some snapshots of lattice occupancy with and without pulling. The initial high density region of agents disperses more quickly in the pulling case. Pulling allows two agents to move at once, so one effect of pulling is to increase the total rate of movement across all the agents. In high density clusters of agents, most of the possible moves are away from the cluster. Therefore pulling has the observed effect of causing high density regions of agents to disperse more quickly  However, we see that the difference between the two cases is less clear as the agents spread out. This is expected because in low density regions it is less likely that two agents will be adjacent, so fewer pulling moves are possible and its influence is diminished. 

In the next section we derive partial differential equations which describe the evolution of the average lattice occupancy. By analysing these equations we can gain some insight into the behaviour observed in Figure \ref{fig:comparison_pulling_vs_non_pulling}.  

\subsection{Continuum model for average occupancy} \label{sec:Cont_model}

To derive the corresponding population-level model, which describes the mean occupancy of the lattice over many repeats, we first introduce some notation. Let $C_r(i,j,t)$ be the occupancy of site $(i,j)$ at time $t$ on the $r^{\text{th}}$ run of the simulation. If we run the simulation $R$ times then the average occupancy of site $(i,j)$ at time $t$ across all simulations is given by 
\begin{equation}
C(i,j,t)=\frac{1}{R}\sum_{r=1}^{R}C_r(i,j,t).
\end{equation}
Now by considering possible changes of occupancy of site $(i,j)$ in a small period of time, $dt$, we can write down the following mean-occupancy equation for the mean occupancy of site $(i,j)$.

\begin{align}
&C(i,j,t+dt)-C(i,j,t)=\nonumber \\ 
&-\frac{p}{4}C(i,j,t)\Big[2(1-C(i+1,j,t))(1-C(i-1,j,t))+2(1-C(i,j+1,t))(1-C(i,j-1,t))\nonumber \\
&+(1-w)\big\{ C(i+1,j,t)(1-C(i-1,j,t))+C(i-1,j,t)(1-C(i+1,j,t))\nonumber \\
&+C(i,j+1,t)(1-C(i,j-1,t))+C(i,j-1,t)(1-C(i,j+1,t))\big\} \nonumber \\
&+w\big\{ C(i+1,j,t)(1-C(i+2,j,t))+C(i-1,j,t)(1-C(i-2,j,t))\nonumber \\
&+C(i,j+1,t)(1-C(i,j+2,t))+C(i,j-1,t)(1-C(i,j-2,t))\big\}\Big]dt\nonumber \\
&+\frac{p}{4}(1-C(i,j,t))\Big[ C(i+1,j,t)+C(i-1,j,t)+C(i,j+1,t)+C(i,j-1,t)\Big]dt.
\label{eq:PMEpull}
\end{align}

There are two mechanisms by which a site can lose occupancy. The first way is that an agent actively moves to a neighbouring site and does not pull an agent along with it. This could be because there is no agent in the pulling position or because there is one in the pulling position but the pull is chosen, with probability $1-w$, to not occur. This is described by the first three lines after the equals sign of equation \eqref{eq:PMEpull}. The second way a site may lose occupancy is that the agent is pulled out of the site by one of its neighbours, this corresponds to the fourth and fifth line after the equals sign of equation \eqref{eq:PMEpull}. There is only one mechanism by which a site can gain occupancy and this is represented by the final line in equation \eqref{eq:PMEpull}. We note here that we have assumed the occupancy of two adjacent sites is independent. The extent to which this assumption is valid has a strong role in determining the agreement between the ABM and the PDE we are about to derive.

\begin{figure}[htbp]
\begin{center} 
\subfigure[]{
\includegraphics[width=0.45\textwidth]{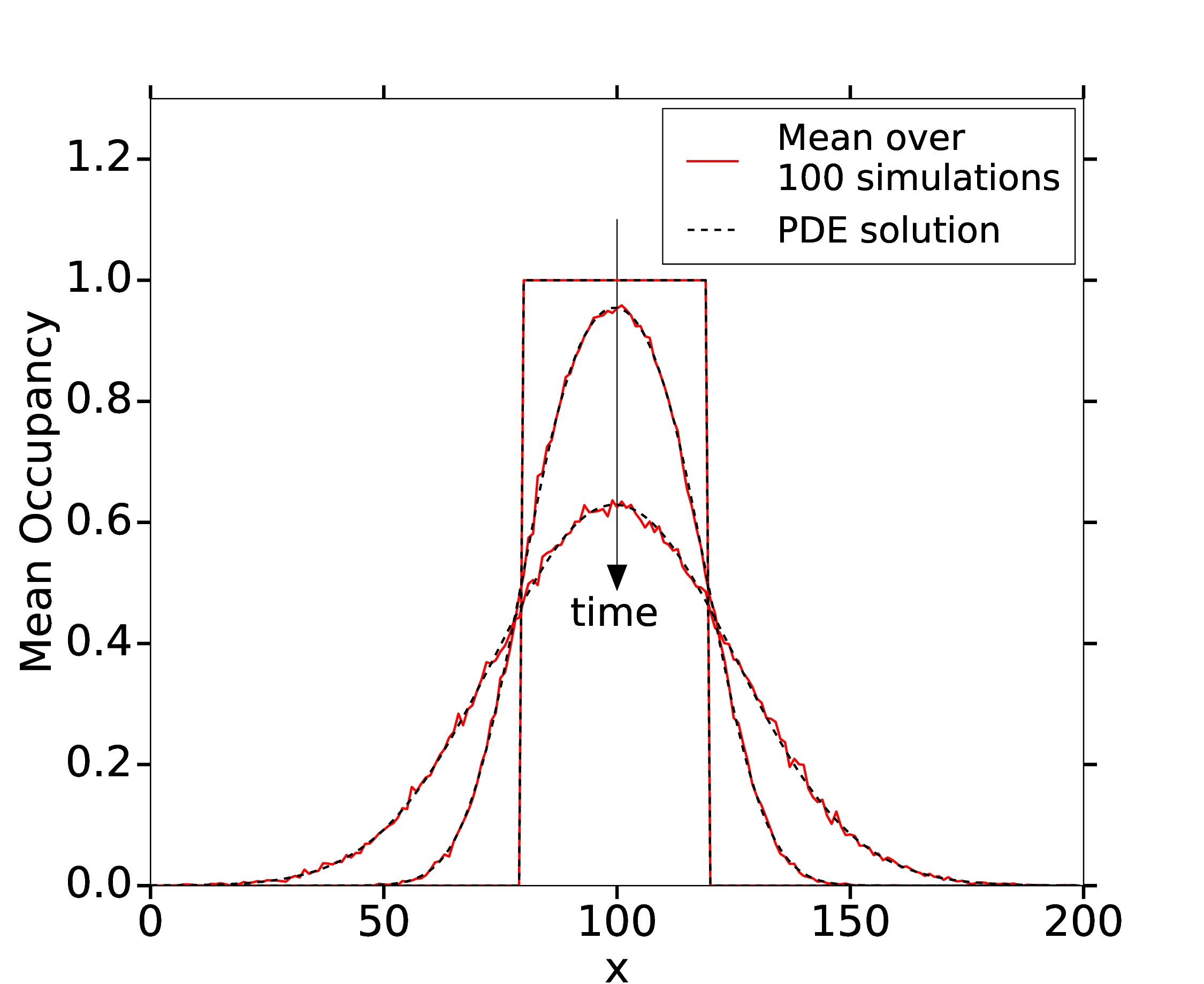}
\label{subfigure:ex_pde_sim}
}
\subfigure[]{
\includegraphics[width=0.45\textwidth]{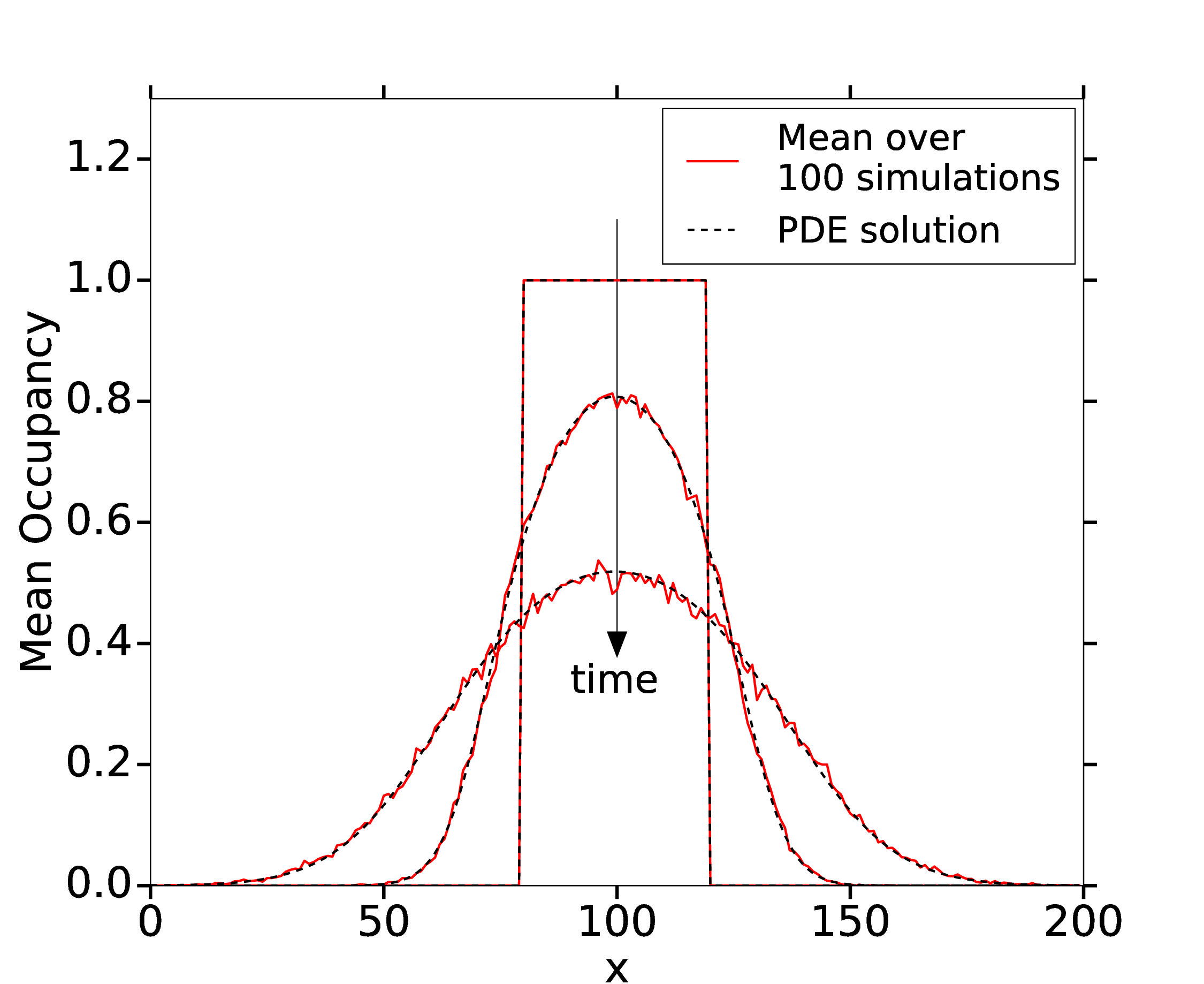}
\label{subfigure:pull_pde_sim}
}
\end{center}
\caption{A comparison between the averaged  behaviour of the ABM and the numerical solution to the corresponding PDE for \subref{subfigure:ex_pde_sim} the simple exclusion process and \subref{subfigure:pull_pde_sim} the exclusion process with cell pulling ($w=1$). All model parameters and initial conditions are the same as for Figure \ref{fig:comparison_pulling_vs_non_pulling}, and the simulations are averaged over 100 repeats. The times shown are $t=0$, $t=200$ and $t=1000$. The dashed black line is the PDE solution, continuous red line is the averaged column density of the ABM, $\bar{C}(i,t)$.} 
\label{fig:pde_ex_pull}
\end{figure}

To obtain a continuum equation we Taylor expand the appropriate terms in equation \eqref{eq:PMEpull} to second-order around site $(i,j)$ and take the limit as the site size, $\Delta$, and the time-step, $dt$, tend to zero, while the ratio $\Delta^2/dt$ remains constant. We obtain the following partial differential equation. 
\begin{equation}
\frac{\partial C}{\partial t}=\nabla \cdot \big[D(1+3wC^2)\nabla C\big]
\label{eq:PDEpull}
\end{equation}
where $$ D=\lim_{\Delta ,dt \rightarrow 0}
\frac{p\Delta^2}{4dt}. $$
With zero-flux boundary conditions at $x=0$ and $x=L_x$ and periodic boundary conditions at $y=0$ and $y=L_y$. We see that the effect of pulling is to enhance the diffusion coefficient with the addition of the term $3wC^2$ resulting in faster dispersion in high density regions. 
In Figure \ref{fig:pde_ex_pull} we compare the averaged column density of the ABMs, given by $\bar{C}(i,t)=1/L_y\sum_{j=1}^{L_y}C(i,j,t)$, with the solution of the one-dimensional version of equation \eqref{eq:PDEpull} obtained by averaging over the $y$-direction both with and without pulling. We see that the agreement between the averaged ABM and the PDE solution is very good and that pulling causes a faster dispersal from the initial high density region. 

\subsection{Comparison with cell pushing} \label{subsec:Comparison_with_pushing}
In this section we compare the population-level effects of the pulling mechanism presented in this paper to the pushing mechanisms presented by \citet{yates2015ipe}. Cell-cell pushing allows active cells to push neighbouring cells out of the way in order to make space for their own moves. In the most basic of the pushing models an agent at position $(i,j)$ which has chosen to move rightward into an occupied site at $(i+1,j)$ can push the agent at $(i+1,j)$ to the right into site $(i+2,j)$, with probability $q$, providing that site is unoccupied. If the site $(i+2,j)$ is occupied then the pushing event is aborted. The mean-occupancy equation for this model is equation \eqref{SUPP-eq:pushing_PME} in the supplementary material.  In Figure \ref{fig:comparison_pushing_vs_pulling}, we present snapshots of the lattice occupancy at various times.

\begin{figure}[htbp]
\begin{center}
\begin{minipage}{.49\textwidth}
\subfigure[]{
\includegraphics[width=1\textwidth]{./snapshot_av_ex_0_py.eps}
\label{subfigure:pulling0_1}
}
\subfigure[]{
\includegraphics[width=1\textwidth]{./snapshot_av_pull_1_200_py.eps}
\label{subfigure:pulling200_1}
}
\subfigure[]{
\includegraphics[width=1\textwidth]{./snapshot_av_pull_1_1000_py.eps}
\label{subfigure:pulling1000_1}
}
\end{minipage}
\begin{minipage}{.49\textwidth}
\subfigure[]{
\includegraphics[width=1\textwidth]{./snapshot_av_ex_0_py.eps}
\label{subfigure:pushing0}
}
\subfigure[]{
\includegraphics[width=1\textwidth]{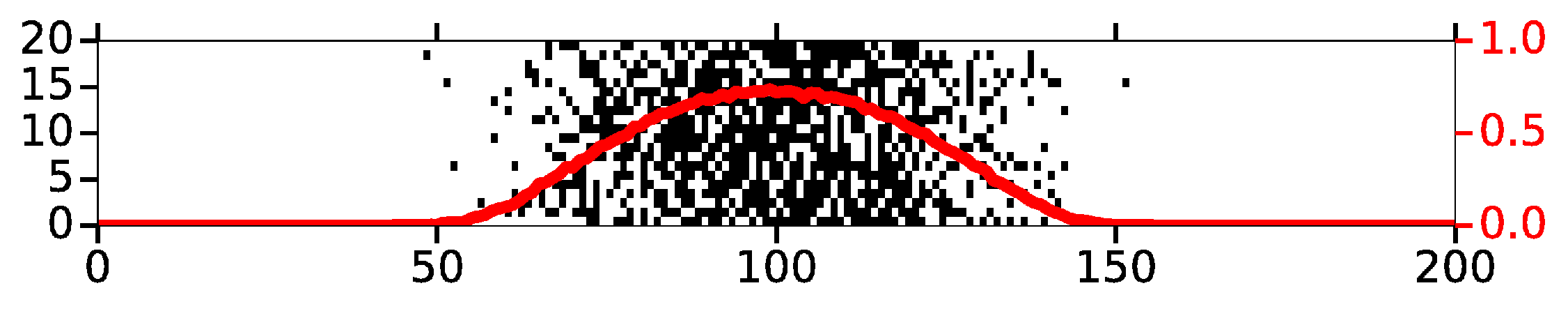}
\label{subfigure:pushing200}
}
\subfigure[]{
\includegraphics[width=1\textwidth]{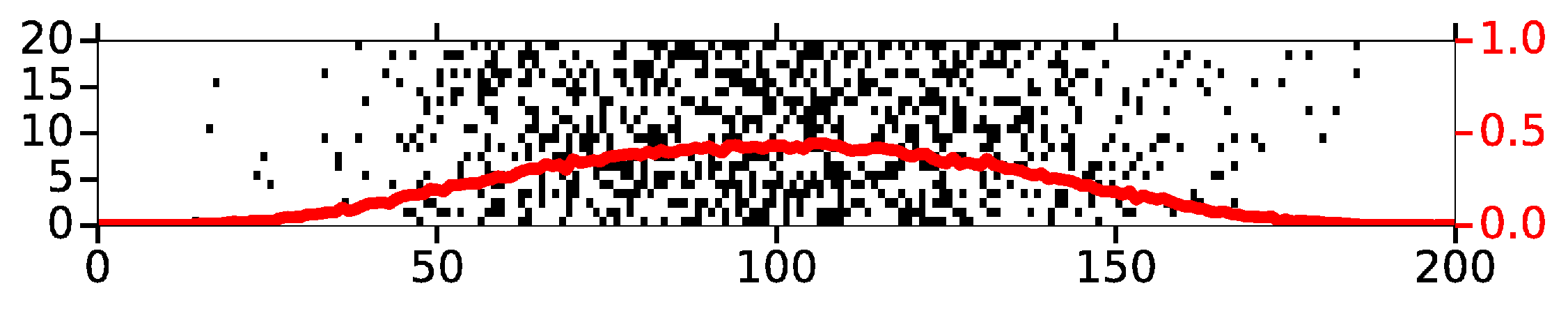}
\label{subfigure:pushing1000}
}
\end{minipage}
\end{center}
\caption{A comparison of pushing and pulling cell migration models. The evolution of the lattice occupancy for \subref{subfigure:pulling0_1}-\subref{subfigure:pulling1000_1} the exclusion process model with simple pushing ($q=1$) and \subref{subfigure:pushing0}-\subref{subfigure:pushing1000} the exclusion process model with agent-agent pulling ($w=1$). The lattice and the initial conditions are the same used in Figure \ref{fig:comparison_pulling_vs_non_pulling}. Lattice occupancy is shown at $t=0$, $t=200$ and $t=1000$. Agents are slightly more spread out in the pushing case than the pulling case.}
\label{fig:comparison_pushing_vs_pulling}
\end{figure}

It is not immediately obvious that these two models will exhibit different behaviours. However we observe in Figure \ref{fig:comparison_pushing_vs_pulling} that the pushing mechanism, in which the pushing probability $q$ is maximal, causes agents to spread out marginally more quickly than the pulling mechanism, when the pulling parameter is maximal. We can explain this phenomenon by considering a high density situation (as in the initial conditions). In the pulling model, only the agents on the outside are able to initiate moves, pulling their neighbours with them. While in the pushing model there is an additional layer of agents behind the outer agents who are also able to initiate moves, pushing their outer neighbours with them. This means the average rate of net movement is higher in the pushing case and consequently high density regions of agents disperse more quickly.

\citet{yates2015ipe} used the same method to derive a partial differential equation from the mean-occupancy equation for simple pushing.  We reproduce the PDE here for comparison. 
\begin{equation}
\frac{\partial C}{\partial t}=\nabla \cdot \big[D(1+4qC)\nabla C\big]
\label{eq:PDEpush}
\end{equation}
where $q$ is the pushing probability.

Comparing equations \eqref{eq:PDEpull} and \eqref{eq:PDEpush} we observe that that, if the pulling probability, $w$, is equal to the pushing probability, $q$, then the effective diffusion constant of the pulling model, $D(1+3wC^2)$, is always smaller than that of the pushing model, $D(1+4qC)$. Also, the density-dependence of the diffusion coefficient for the pulling model is quadratic, so its effect decreases more rapidly with decreasing density than the density-dependent diffusion coefficient associated with pushing, which is only linear. This is consistent with the behaviour of the ABMs we observed in Figure \ref{fig:comparison_pushing_vs_pulling}. 

The differences in the density-dependencies of the diffusion coefficients between the two models can be understood by considering the two mean-occupancy equations. In the case of pulling the quadratic terms cancel, whereas in the pushing mean-occupancy equation the cubic terms cancel. The cancellation of the cubic terms in the pushing PME can be interpreted as follows. Averaged over many repeats, the probability of an agent being pushed into a previously vacant site is equal to the probability of an agent pushing its way out the site it occupies. We do not have the same symmetry in the pulling case since it is not possible for an agent to be pulled into a previously vacant site. We do however have a different symmetry. In one way the presence of pulling decreases the probability of a site losing occupancy because it increases the probability of an adjacent agent being dragged into the site when the agent initially occupying the site leaves. In another way it increases the probability of a site losing occupancy because an agent may be pulled out of the site. The balance between these mechanisms causes the quadratic terms to cancel. 
 
\section{Extensions to the pulling mechanism}\label{section:extensions_to_the_pulling_paradigm}
In this section we present some extensions to the simple pulling model that incorporate more complexity. In subsection \ref{subsec:multiple_agents} we consider two types of pulling in which multiple agents can be moved by a single active agent. We combine simple cell-cell pulling with cell-cell pushing in section \ref{subsec:pushing_and_pulling} and give cells the ability to pull at a distance in subsection \ref{subsec:pulling_distance}. We draw together all of the on-lattice pulling mechanisms in subsection \ref{subsection:summary_of_pulling_models}, comparing the macroscopic models they give rise to as well as making direct comparisons between the averaged density of the ABM and the PDE for several of the mechanisms. In subsection \ref{section:error_comparison} we give a more quantitative and dynamic comparison between the averaged ABMs and the solutions to the PDEs we have derived using the histogram distance error (HDE) metric.

\subsection{Pulling multiple agents} \label{subsec:multiple_agents}
A natural extension to the simple pulling model is to allow the active agent to pull multiple agents along with it. There are two slightly different ways of implementing a multiple pulling model, we refer to them as type 1 and type 2 multiple pulling. We describe them below. 

\subsubsection{Type 1 multiple pulling} \label{subsec:multiple_agents_type1}
Given that an agent selected to move is in contact with a neighbour in a position which would allow it to be pulled, the type 1 multiple pulling model specifies a constant probability, $w$, that the first agents pulls one or more agents, independent of how many agents could potentially be pulled. For example, consider the case in which we restrict the active agent to pulling at most two agents. We refer to this as the second-order case. Suppose an agent in site $(i,j)$ has been selected to move rightwards to $(i+1,j)$. If there is an agent in site $(i-1,j)$ then a pull occurs with probability $w$. If a pull is to occur, we next decide how many agents will be pulled. If there is no agent in site $(i-2,j)$ then a single agent must be pulled. However, if there is an agent occupying site $(i-2,j)$ (see Figure \ref{fig:2nd_order_pulling_2_schematic}) then a double-pull occurs with probability $w_1$ and a single pull occurs with probability $(1-w_1)$. In the second-order case we stop here, but for a third-order pulling model we would continue the process by checking if there is agent in site $(i-3,j)$. If there is, given a double-pull has already been chosen then either a triple pull occurs with probability $w_2$ or the agent in site $(i-3,j)$ is ignored with probability $1-w_2$. If a double-pull has not been chosen, then a triple pull is not possible, even if a third agent is in the correct position. The parameter $w_1$ can be interpreted as the probability of a double-pull occurring given that a pull is happening and that there are agents in the appropriate positions. A natural choice might be to set $w_1=w$, so that the probability of a double move is $w^2$. 

The mean-occupancy equation for this model is included as equation (\ref{SUPP-eq:pulling_2nd_1_PME}) in the supplementary material. Using the same method as we used to derive equation \eqref{eq:PDEpull} we obtain the following PDE for second-order type 1 multiple pulling:
\begin{equation}
\frac{\partial C}{\partial t}=\nabla \cdot \Big(D\big(1+3w(1-w_1)C^2+8ww_1C^3\big)\nabla C\Big).
\label{eq:PDEpull_2nd_1}
\end{equation}
We note that the coefficient of the term associated with pairwise movement ($C^2$) is reduced in comparison to the simple pulling PDE. This is because there are now fewer single pulls occurring. There is also an additional cubic term, which is a consequence of the fact that triplets of agents can now move simultaneously in a double-pull move. 

For general $n^{\text{th}}$-order pulling we need parameters $w,w_1,....,w_{n-1}$. The mean-occupancy equations become very lengthy. The corresponding PDE for $n^{\text{th}}$-order pulling is included in Table \ref{Table:PDE_coefficients}.

\subsubsection{Type 2 multiple pulling}
\begin{figure}[h]
\begin{center} . 
\includegraphics[height=0.4\textheight]{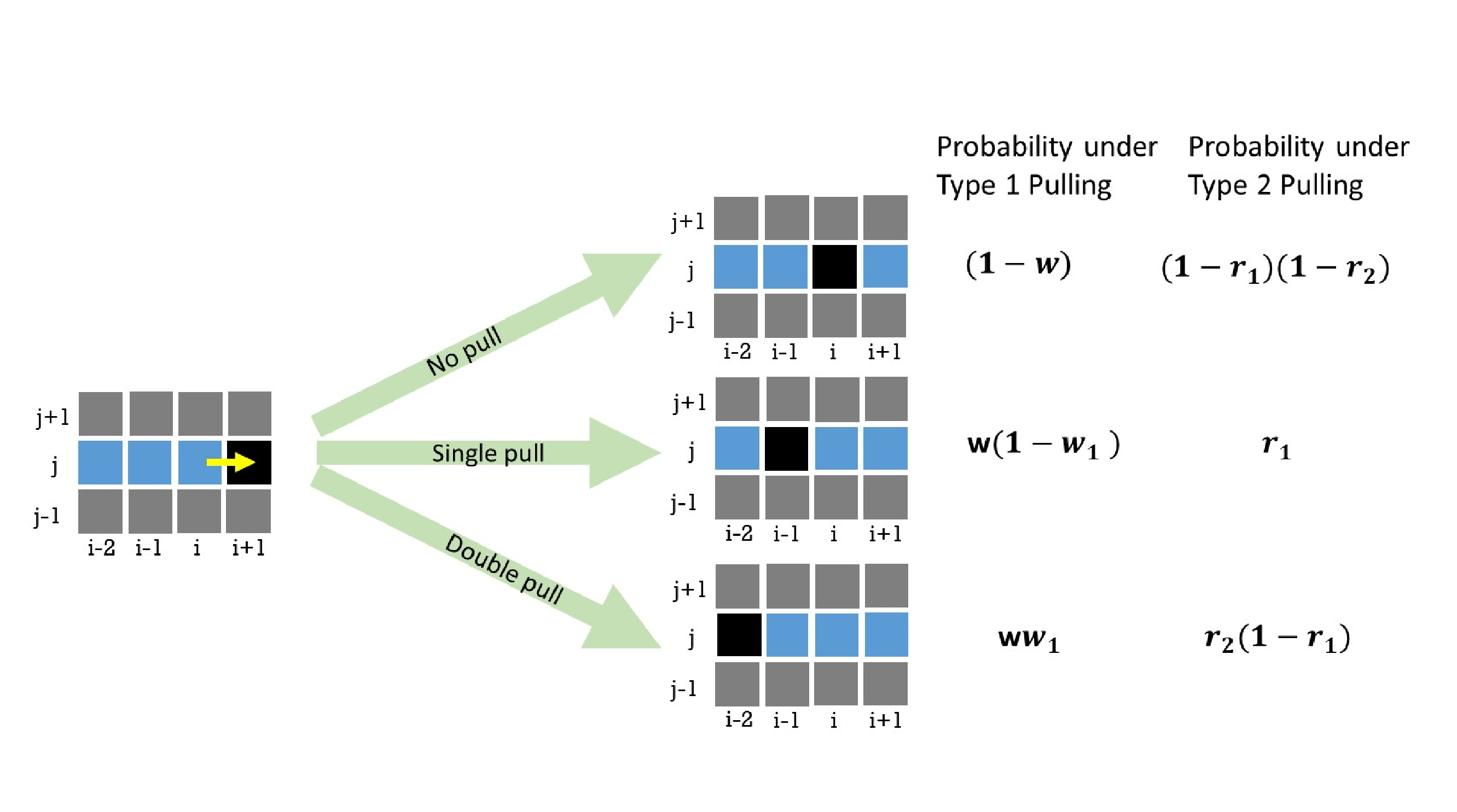}
\end{center}
\caption{Second-order pulling schematic, illustrates type 1 and type 2 multiple pulling. Agents are shown in blue, vacant sites are black. Sites for which the occupancy is not important are grey. The agent in site $(i,j)$ has been selected to move to the right, and sites $(i-1,j)$ and $(i-2,j)$ are both occupied. There are three possible outcomes indicated by the light green arrows. The probabilities of each outcome under the type 1 and type 2 models are shown on the right. An active move is indicated by a yellow arrow.}
\label{fig:2nd_order_pulling_2_schematic}
\end{figure}

There is an alternative way of incorporating multiple agent pulling. To distinguish it from type 1 pulling we again explain a second-order pulling-event. Suppose an agent in site $(i,j)$ is chosen to move rightwards to site $(i+1,j)$. We first check whether there is an agent in site $(i-1,j)$. If there is then we allow a single pull to occur with probability $r_1$. If the single pull is chosen not to occur then we check whether there is an agent in site $(i-2,j)$, if there is then we allow a double-pull to occur with probability $r_2$ (see Figure \ref{fig:2nd_order_pulling_2_schematic}). In this scenario, a second-order pulling event can occur only if a first-order pulling event is chosen not to happen.

Again, we include the mean-occupancy equation \eqref{SUPP-eq:pulling_2nd_2_PME} in the supplementary material. We obtain the following PDE for type 2 second-order pulling.
\begin{equation}
\frac{\partial C}{\partial t}=\nabla \cdot \Big(D\big(1+3r_1C^2+8r_2(1-r_1)C^3\big)\nabla C\Big).
\label{eq:PDEpull_2nd_2}
\end{equation}
This is similar to the type 1 pulling PDE. Indeed, the type 1 and type 2 PDEs are simple re-parameterisations of each other. Interestingly, however, the agent-based models not a simple re-parametrisations of each other since the distribution of first and second-order jumps are different between the two models.

\subsection{Combined pushing and pulling}\label{subsec:pushing_and_pulling}

Another extension we make to the simple pulling model (i.e. first-order pulling only) is to combine it with the simple pushing model presented in \citet{yates2015ipe} and described in section \ref{subsec:Comparison_with_pushing} of this manuscript.

Suppose an agent in site $(i,j)$ is chosen to move right to site $(i+1,j)$. If there is no agent in site $(i+1,j)$ then the move is completed. If there is an agent in site $(i-1,j)$ then the active agent pulls this second agent with it with probability $w$. If there is an agent in site $(i+1,j)$ but not in site $(i+2,j)$ the the active agent pushes the agent at $(i+2,j)$ with probability $q$. If the push occurs and there is an agent in site $(i-1,j)$ then a pull also happens with probability $w$.

Using the mean-occupancy equation \eqref{SUPP-eq:pulling+pushing_PME} of the supplementary material and following the same procedure as for equations \eqref{eq:PDEpull}, \eqref{eq:PDEpull_2nd_1} and \eqref{eq:PDEpull_2nd_2}, we obtain the following PDE
\begin{equation}
\frac{\partial C}{\partial t}=\nabla \cdot \Big(D\big(1+4qC+3w(1-q)C^2+8qwC^3\big)\nabla C\Big).
\label{eq:PDEpull+push}
\end{equation}
This is very similar to the type 1, second-order pulling PDE (equation \eqref{eq:PDEpull_2nd_1}) with $q$, the pushing probability playing the role of $w_1$, the probability of a double-pull occurring given that a first cell has already been pulled. This similarity is unsurprising since both the double-pull and the push-and-pull movements corresponding to terms $3w(1-w_1)C^2$ and $3w(1-q)C^2$ respectively in equations \eqref{eq:PDEpull_2nd_1} and \eqref{eq:PDEpull+push}, respectively, represent the movement of three consecutively aligned agents.  The additional  term ($4qC$) in equation \eqref{eq:PDEpull+push} corresponds to pure pushing events, in which no pulling occurs, and is the same as the term by which diffusion in enhanced in the pure pushing PDE given by equation \eqref{eq:PDEpush}. 

\subsection{Pulling at a distance} \label{subsec:pulling_distance}
\begin{figure}[htpb]
\begin{center} 
\includegraphics[width=\textwidth/3]{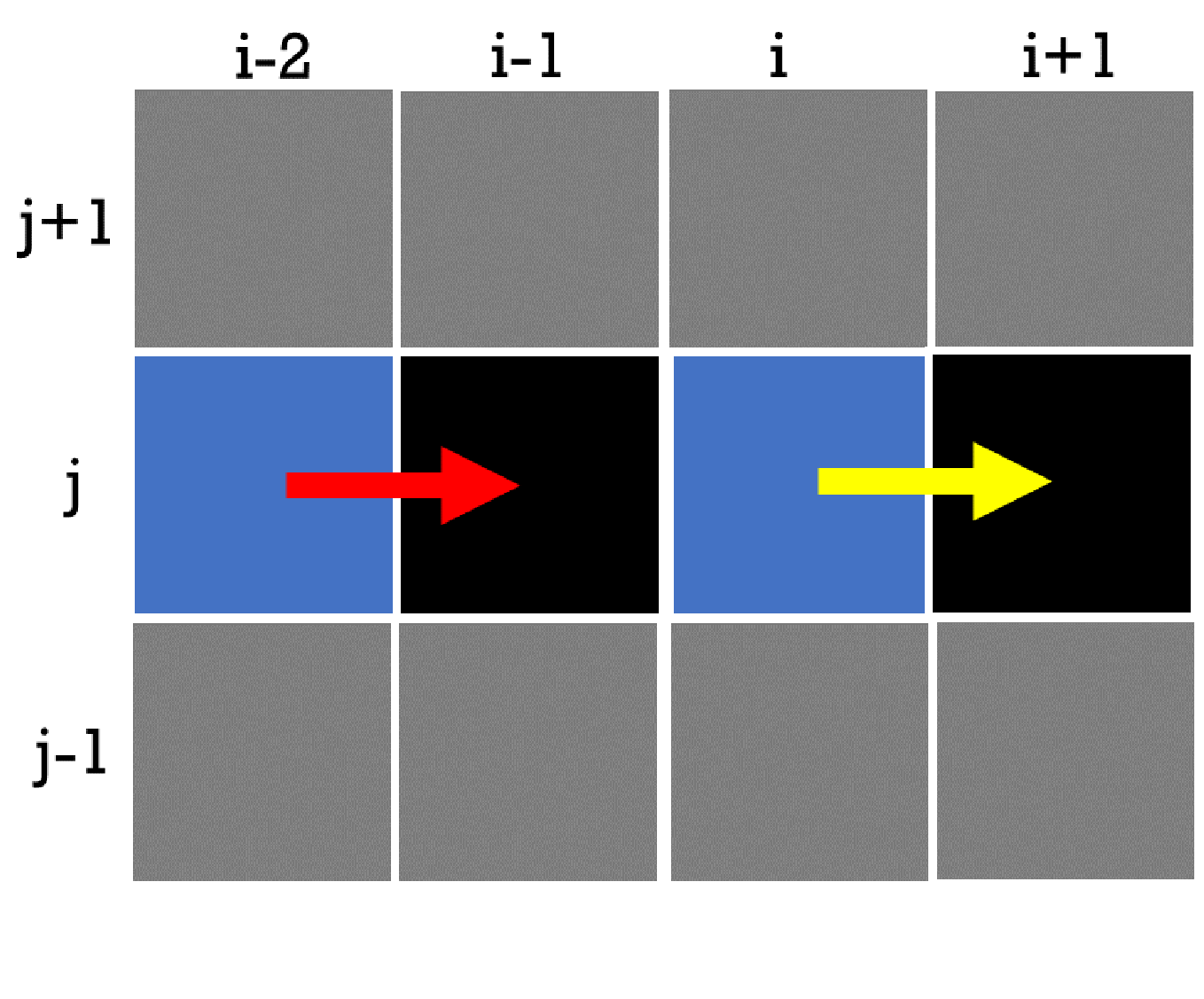}
\end{center}
\caption{This shows a pulling-at-a-distance event. The colour scheme is the same as for Figure \ref{fig:cell_schematics}. The agent in site $(i,j)$ has been selected to move rightwards. The site $(i-1,j)$ is vacant but there is an agent present in site $(i-2,j)$. The agent in site $(i,j)$ moves into site $(i+1,j)$ and pulls the agent in site $(i-2,j)$ into site $(i-1,j)$. }
\label{fig:pulling_distance schematic}
\end{figure}

Cells are able to interact without the majority of their cell membranes being in contact with each other by extending filopodia \cite{kulesa1998ncc, kasemeier2005inc}. It therefore seems reasonable to consider a case in which the active agent is able to pull other agents which are nearby, but not necessarily directly neighbouring. For illustrative purposes we consider the simplest such model. Suppose an agent in position $(i,j)$ is chosen to move rightwards to position $(i+1,j)$. If there is agent in position $(i-1,j)$ then, as before, this agent is pulled into position $(i,j)$ with probability $w$. If there is not an agent in position $(i-1,j)$ then we next check whether there is agent in position $(i-2,j)$. If there is an agent in position $(i-2,j)$ then, with probability $v$, this agent is pulled into site $(i-1,j)$ maintaining its distance from the pulling agent. This event is depicted schematically in Figure \ref{fig:pulling_distance schematic}. We derive the following PDE for the mean occupancy of position $x$ at time $t$ in the usual manner:
\begin{equation}
\frac{\partial C}{\partial t}=\nabla \cdot \Big(D\big(1+3wC^2-v(2C-10C^2+8C^3)\big)\nabla C\Big).
\label{eq:PDEpull_d}
\end{equation}
We note that there is then a critical value of density $C^*=1/4$, for which, if $C<C^*$, the pulling-at-a-distance mechanism has the effect of reducing the diffusion coefficient so agents disperse more slowly that they would with simple nearest-neighbour pulling. For $C>C^*$ the mechanism enhances the diffusion coefficient.

\subsection{Summary of pulling models}\label{subsection:summary_of_pulling_models}
For all the models we have considered thus far, we have derived a non-linear diffusion equation from the mean-occupancy equation in the limit as the site size, $\Delta$, and time-step, $dt$, tend to zero while the ratio $\Delta^2/dt$ remains constant. These PDEs all have the following form
\begin{equation*}
\frac{\partial C}{\partial t}=\nabla \cdot \Big(D(C)\nabla C\Big).
\end{equation*}
Where $D(C)$ is the density-dependent diffusion coefficient. Table \ref{Table:PDE_coefficients} summarises these coefficients for each of the variants of the pulling model. 
\begin{center}
  \begin{table}[h!!!!]
    \setlength{\tabcolsep}{16pt}
    \small
    {\renewcommand{\arraystretch}{1.25}
    \begin{tabular}{|c|c|} \hline
      Pulling type  & Diffusion coefficient  \\ \hline
      Simple pulling  & $D(1+3wC^2)$   \\  \hline
      Type 1, second-order pulling  & $D\left(1 + 3w(1-w_1)C^2+8ww_1C^3\right)$   \\ \hline
      Type 1, $n^{\text{th}}$-order pulling & $ D\left(1+w\sum_{i=2}^{n+1}(i^2-1)\prod_{k=1}^{i-2}w_k\prod_{k=i-1}^{n-1}(1-w_k)C^i\right) $\\ \hline
      Type 2, second-order pulling   & $D\left(1 + 3r_1C^2+8r_2(1-r_1)C^3\right)$   \\ \hline
      Pushing and pulling  & $D\left(1+4qC+3q(1-w)C^2+8qwC^3\right)$   \\ \hline
      Pulling at a distance  & $D\left(1+3wC^2-v(2C-10C^2+8C^3)\right)$   \\ \hline
    \end{tabular}
    \caption{Different forms of the density-dependent diffusion coefficient for the various pulling models.}
    \label{Table:PDE_coefficients}
    }
    \end{table}
  \end{center}


\begin{figure}[h!!!!]
\begin{center} 
\subfigure[] {
\includegraphics[width=.45\textwidth]{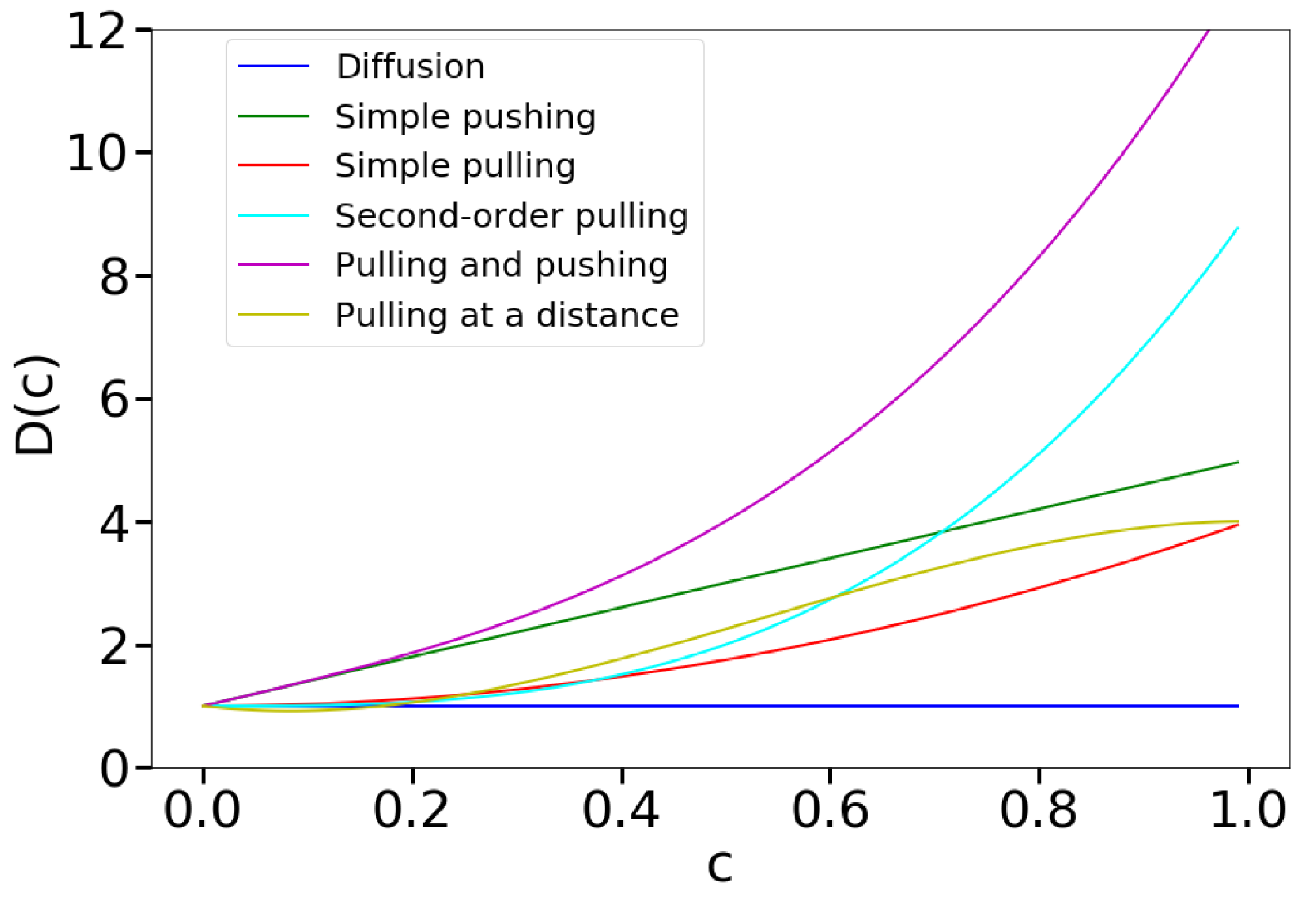}
\label{fig:diff_coeffs}
}
\subfigure[]{
\includegraphics[width=.45\textwidth]{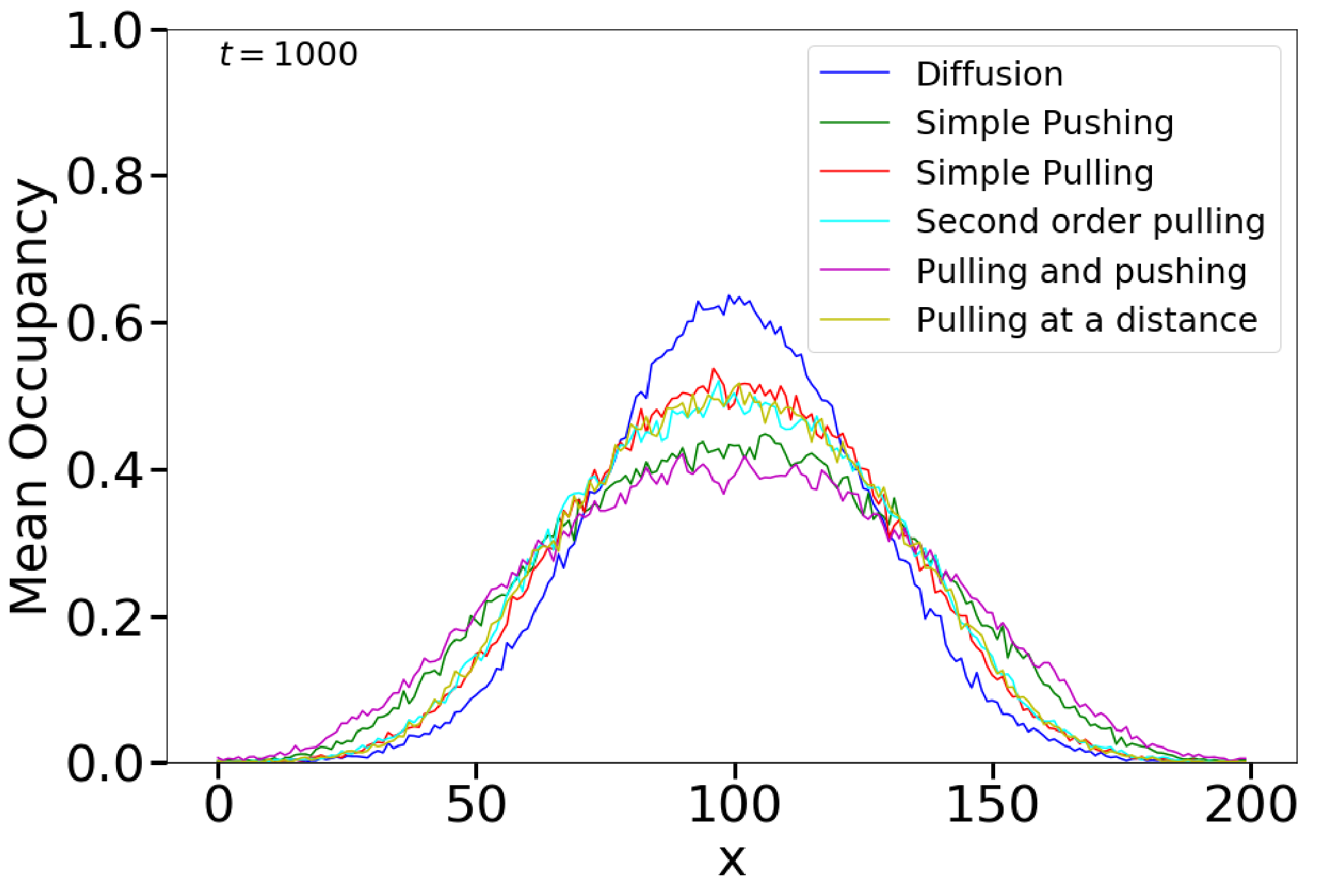}
\label{fig:onlat_abm_comp}
}
\end{center}
\caption{Panel \subref{fig:diff_coeffs} illustrates the dependence of the diffusion coefficient on density, for the various pulling models. All pulling and pushing parameters are set to unity and $p$, the attempted jump rate, is chosen so that $D=1$. Panel \subref{fig:onlat_abm_comp} shows the averaged column density of the ABMs with 100 repeats at $t=1000$. All simulations are carried out on the same lattice with the same initial and boundary conditions as in Figure \ref{fig:comparison_pulling_vs_non_pulling} }
\label{fig:abm_comp}
\end{figure}

In Figure \ref{fig:abm_comp} we compare the population level effects of the various pulling mechanisms. Figure \ref{fig:abm_comp} \subref{fig:diff_coeffs} contrasts the derived density-dependent diffusion coefficients of the continuum models, while Figure \ref{fig:abm_comp} \subref{fig:onlat_abm_comp} compares the averaged density profiles of the various ABMs at $t=1000$. Unsurprisingly, the models with the largest diffusion coefficients cause the agents to spread out more quickly. For example the mechanism combining pulling and pushing causes agents to disperse most quickly and has the largest diffusion coefficient at all values of $c$. An initially surprising feature is that the simple pulling density profile is very similar to the second-order pulling density profile and that of the pulling-at-a-distance mechanism, despite the the models having seemingly quite different dependence of their diffusion coefficients on density. This is because the diffusion coefficient for the second-order pulling mechanism is significantly different from that of the other two mechanisms only when density is high. In turn this difference can be traced to the agent-based model. It is only at high densities, in the agent-based model, that double-pulls occur with any frequency. Agents can spread out more quickly in high-density regimes (such as provided that by the initial condition), but as the density decreases any spreading advantage is quickly eroded.
  
\begin{figure}[b]
\begin{center}
\begin{minipage}{.49\textwidth}
\subfigure[]{
\includegraphics[width=.9\textwidth]{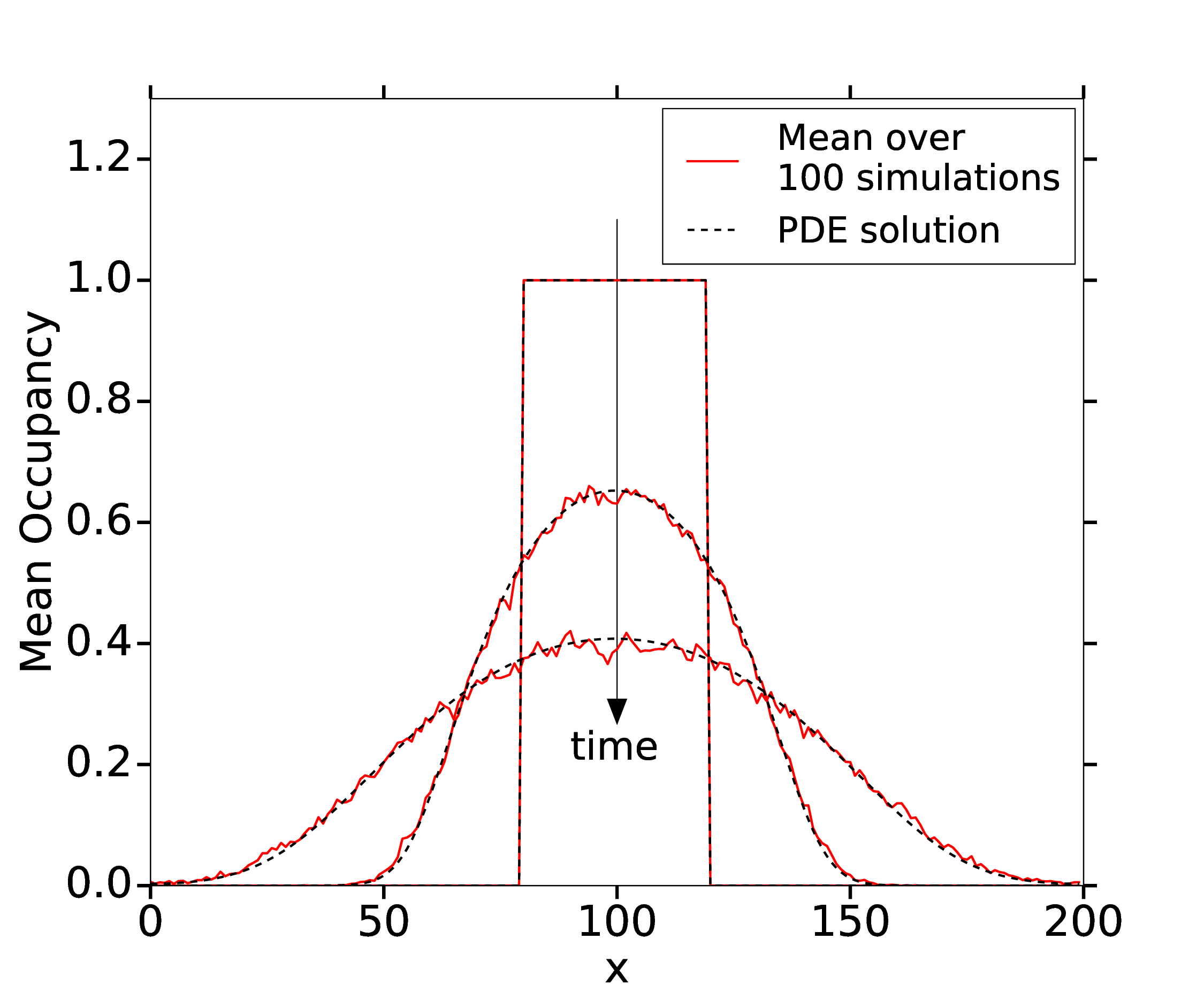}
\label{subfigure:pullpush_sim}
}
\subfigure[]{
\includegraphics[width=.9\textwidth]{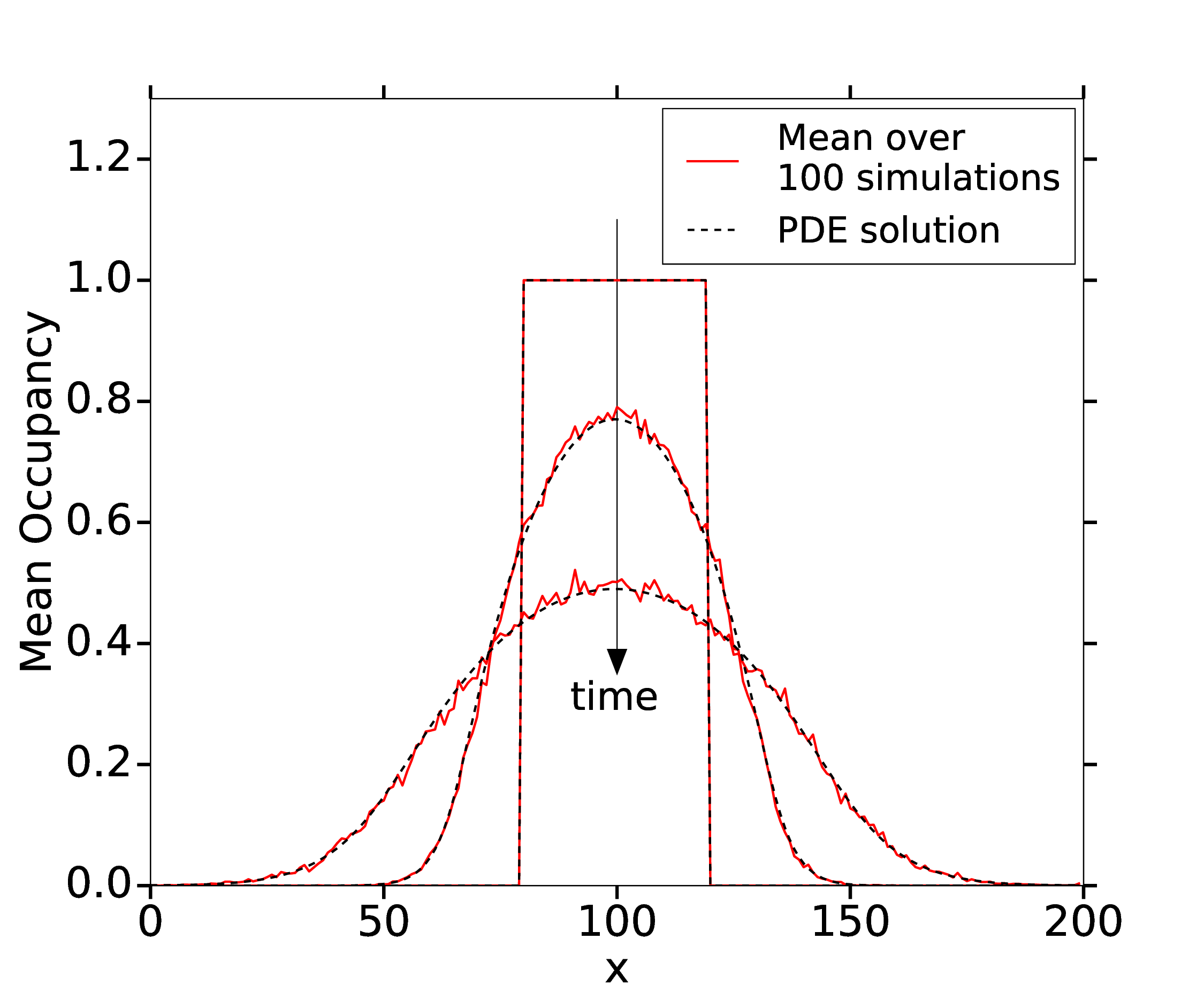}
\label{subfigure:pull_dist_sim}
}
\end{minipage}
\begin{minipage}{.49\textwidth}
\subfigure[]{
\includegraphics[width=.9\textwidth]{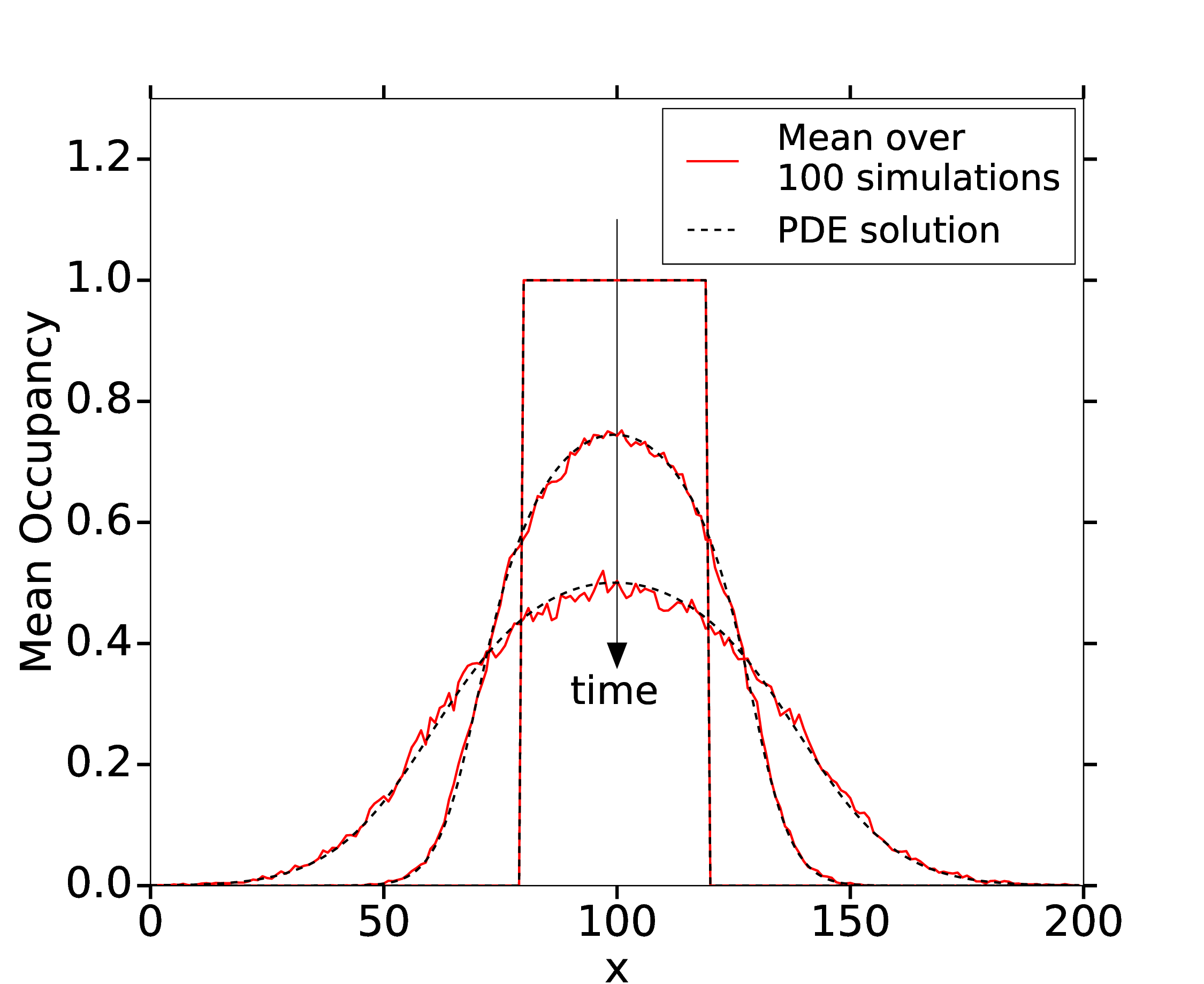}
\label{subfigure:pull_2nd_sim}
}
\subfigure[]{
\includegraphics[width=.9\textwidth]{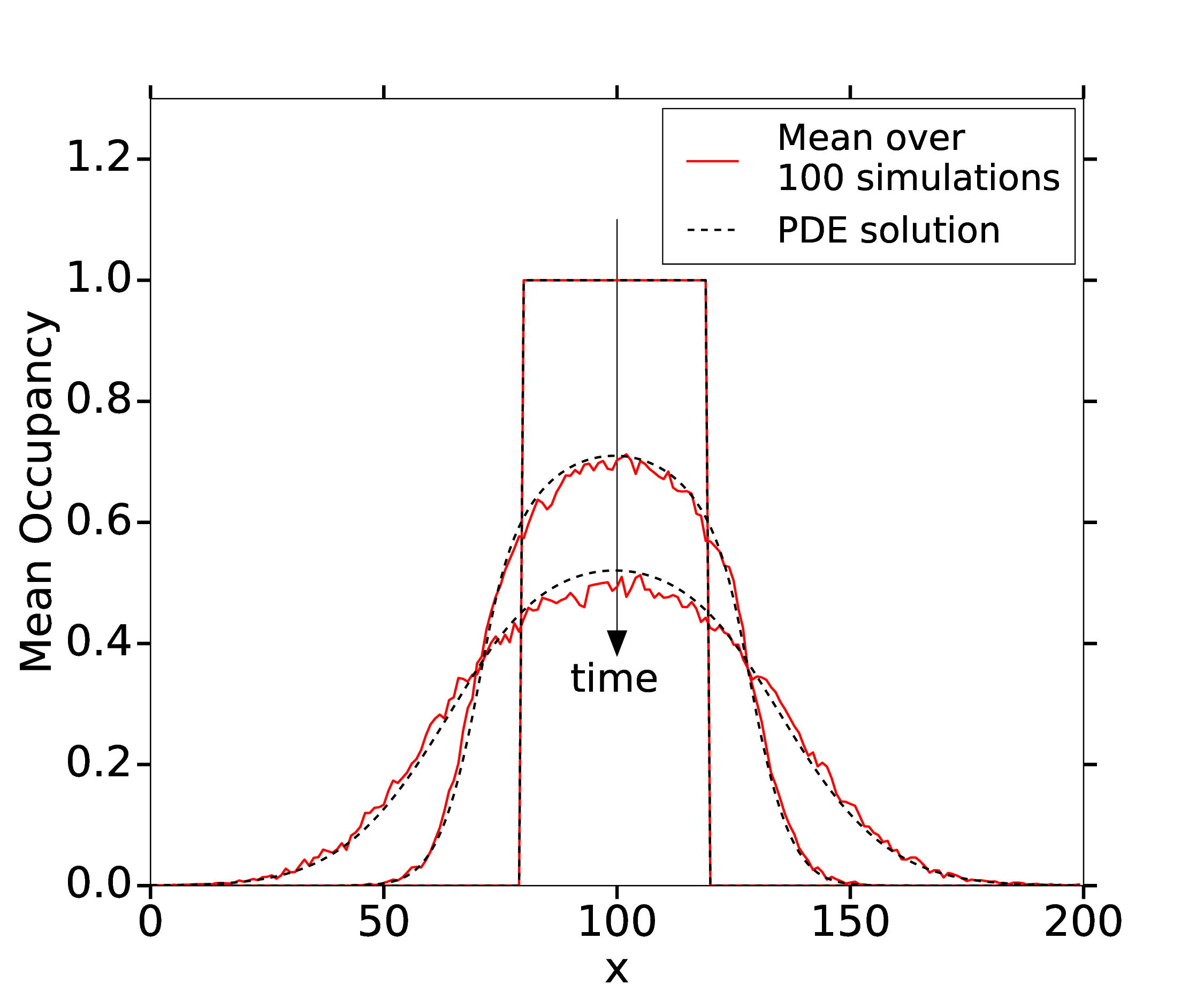}
\label{subfigure:pull_5th_sim}
}
\end{minipage}
\end{center}
\caption{A comparison between the averaged column density and the corresponding PDE for the various extensions of the pulling models. Panel \subref{subfigure:pullpush_sim} shows the comparison between the ABM combining pulling with pushing and its corresponding mean-field PDE given by equation \eqref{eq:PDEpull+push} (see section \ref{subsec:pushing_and_pulling}) with $q=w=1$. Panel \subref{subfigure:pull_dist_sim} is for the pulling-at-a-distance model (see section \ref{subsec:pulling_distance}) with $w=1$. \subref{subfigure:pull_2nd_sim} is for the type 1, second-order pulling model (see section \ref{subsec:multiple_agents_type1}) with $w=w_1=1$. \subref{subfigure:pull_5th_sim} is for the type 1, $5^{\text{th}}$-order pulling model (see section \ref{subsec:multiple_agents_type1}) with $w=w_1=w_2=w_3=w_4=1$. Simulations are carried out on the same lattice with the same initial and boundary conditions as in Figure \ref{fig:comparison_pulling_vs_non_pulling}}
\label{fig:sim_vs_PDE}
\end{figure}

In Figure \ref{fig:sim_vs_PDE} we compare the agreement between the averaged column density ABM and the numerical solution of the corresponding PDE for several of the pulling models presented in this section. We can see that the fit remains very good in most cases. However, in the case of $5^{\text{th}}$-order pulling (see Figure \ref{fig:sim_vs_PDE} \subref{subfigure:pull_5th_sim}) the agreement appears weaker. In the next section we quantify the error between the PDE and ABM for the various models, and explain the discrepancies.

\subsection{Error comparison}\label{section:error_comparison}
In order to quantify the error between the averaged ABMs and the solutions to the corresponding PDEs we use the histogram distance error (HDE)
\begin{equation*}
H(t)=\sum_{i=1}^{L_x}\frac{\mid a_i(t)-b_i(t) \mid}{2}.
\end{equation*}
Here $a_i$ is the normalised, averaged density of column $i$, obtained by repeat simulations of the ABM, and $b_i$ is the solution to the corresponding normalised PDE at the mesh point corresponding to the centre of column $i$. The HDE is normalised so that it takes a maximum value of unity when the two histograms have disjoint supports. 

\begin{figure}[]
\begin{center} 
\includegraphics[width=\textwidth]{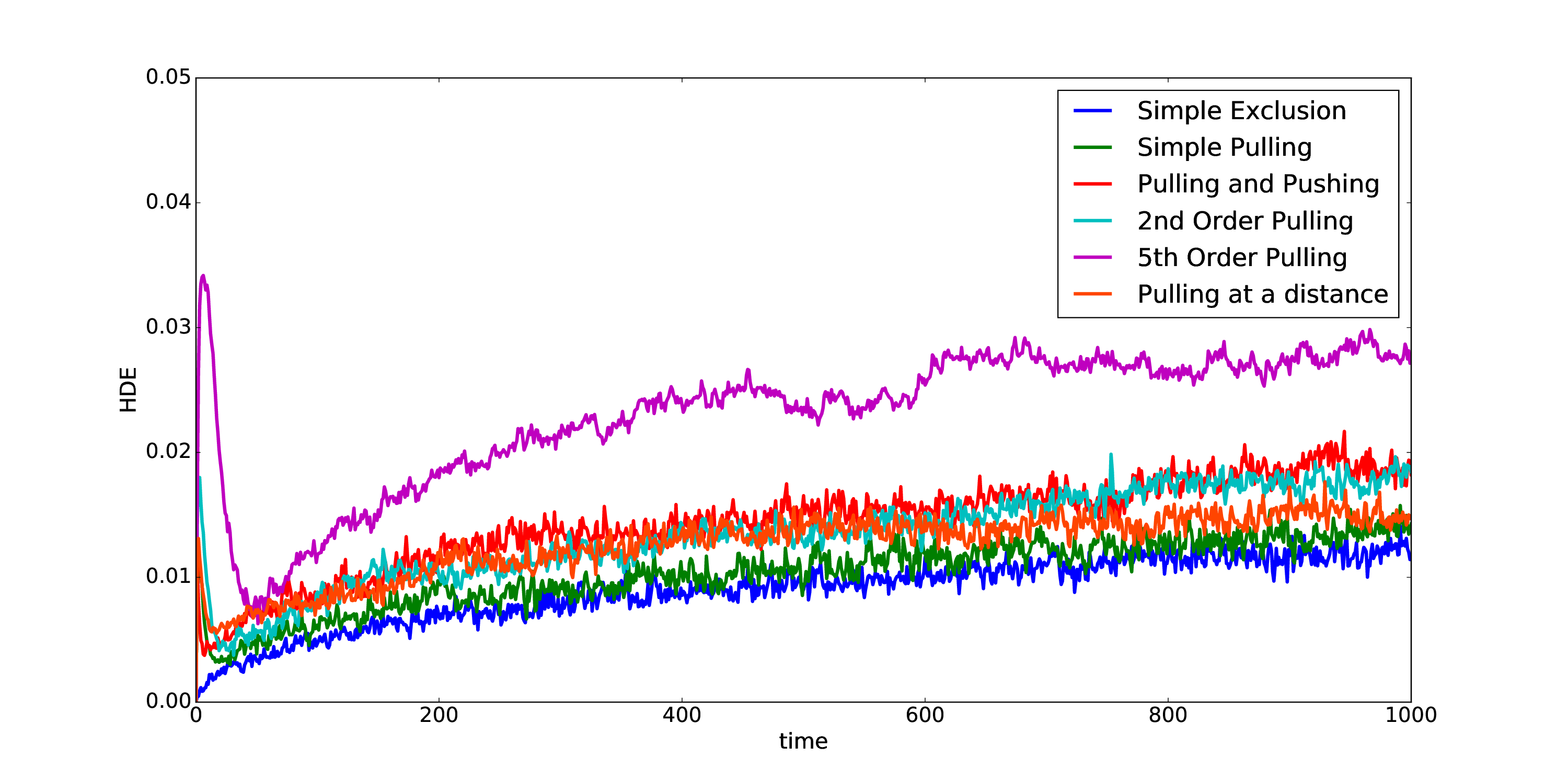}
\end{center}
\caption{A comparison of the HDE between the mean-field PDE solution and the averaged ABM density for some of the different models. We use the same lattice with the same initial and boundary conditions as for Figures \ref{fig:pde_ex_pull} and \ref{fig:sim_vs_PDE}. All pulling probabilities are unity as they are in Figure \ref{fig:pde_ex_pull} and Figure \ref{fig:sim_vs_PDE}. We see that the error is smallest for the simple exclusion model and largest for the $5^{\text{th}}$-order pulling model.}
\label{fig:hde_comparison}
\end{figure}

Figure \ref{fig:hde_comparison} compares the evolution of the HDE for some of the different models. We see that the error is fairly small for all models but, as expected, the fit between the simple exclusion process and its corresponding PDE is the best, and the fit between the $5^{\text{th}}$-order pulling model and its corresponding PDE is the worst. The assumption of independence in lattice occupancy used for deriving the PDEs becomes progressively less valid the more sites are involved in the agent-agent interactions.

\section{Off-lattice models} \label{chapter:Off lattice models}
In this section we investigate the extent to which the lattice structure we previously imposed on our ABMs alters the models' behaviours. To do this we consider off-lattice models in which agents are free to move on a continuous domain. For simplicity we only study the one-dimensional case. For each ABM that we specify, we derive a corresponding population-level model in the continuum limit. In subsection \ref{sec:Model_1} we consider a model in which agent moves that would result in an overlap are aborted. Contrastingly in subsection \ref{sec:Model_2} we consider a more realistic model in which cells are allowed to make contact with each other. In subsection \ref{subsec:offlat_pulling_distance} we consider a model in which cells can pull each other at a distance. Finally, in subsection \ref{subsection:off_lattice_error_comparison}, we compare the behaviour of the ABMs to the their corresponding population-level descriptions both qualitatively and quantitatively.

In each of the following models $N$ agents, with radius $R$, are initialized on the interval $[0,L]$. An agent whose centre is at position $x$  therefore occupies the interval $[x-R,x+R]\subset [0,L]$.  Since the entirety of each agent must be within $[0,L]$ agents must have their centres in the domain $[R,L-R]$. Agents are selected to move at random, with movement rate $p$, so that $pdt$ is the probability that an agent, chosen uniformly at random, attempts a move in the time interval $[t,t+dt]$. Once an agent, at position $x$, has been selected to move, the movement direction (left or right) is chosen at random each with probability $1/2$. The agent then jumps a distance $d$ so that it now occupies $[x-R+d,x+R+d]$ (if a rightwards move was selected). If a proposed move would result in two agents overlapping then we have two options: the first option is that the move is simply aborted. The second option is the the moving agent moves as far as possible in the proposed direction, so that it ends up in contact with the blocking agent. 

\subsection{Aborting overlapping moves} \label{sec:Model_1}
We first study the case in which we abort moves which would result in an overlap. Let $C_i(x,t)$ be the probability distribution for the centre of agent $i$. We have the following mean-occupancy equation. 
\begin{align}
C_i(x,t+dt)&=C_i(x,t)+\frac{p dt}{2}C_i(x-d,t)\Big(1-\sum_{j\neq i}\int_{2R}^{2R+d}C_j(x-d+s,t)ds\Big) \nonumber \\
&+\frac{p dt}{2}C_i(x+d,t)\Big(1-\sum_{j\neq i}\int_{-2R-d}^{-2R}C_j(x+d+s,t)ds\Big) \nonumber \\
&-\frac{p dt}{2}C_i(x,t)\Big(1-\sum_{j\neq i}\int_{2R}^{2R+d}C_j(x+s,t)ds\Big) \nonumber \\
&-\frac{p dt}{2}C_i(x,t)\Big(1-\sum_{j\neq i}\int_{-2R-d}^{-2R}C_j(x+s,t)ds\Big).
\label{eq:Model_1_PME}
\end{align}
This equation represents the four ways in which the occupancy at position $x$ of agent $i$ can change in the time interval $[t,t+dt]$. The first term on the right hand side describes gaining occupancy in the following way. Agent $i$ was centred at position $x-d$ at time $t$ and was chosen to move rightwards (with probability $p dt/2$). This move can happen as long as there is no part of any other agent occupying the region $[x-d+R,x+R]$ (i.e. no other agents are centred in the region $[x-d+2R,x+2R]$). The second term represents the equivalent leftwards move, i.e. agent $i$ was initially centred at $x+d$. 

The third term describes losing occupancy at position $x$ in the following way. Suppose agent $i$ was centred at position $x$ at time $t$ and was chosen to move rightwards (with probability $p dt/2$). This move is not aborted as long as there is no part of any other agent occupying the region $[x+R,x+d+R]$ (i.e. no other agents are centred in the region $[x+2R,x+d+2R]$). Finally, the fourth represents the loss occupancy caused by the equivalent leftwards move. 

Just as in the on-lattice case, by Taylor expanding and taking appropriate limits we can derive a corresponding partial differential equation. We follow the same argument as \citet{dyson2012mli} to derive the PDE, but first introduce some useful notation. 

\begin{figure}[htbp]
\begin{center} 
\includegraphics[width=0.5\textwidth]{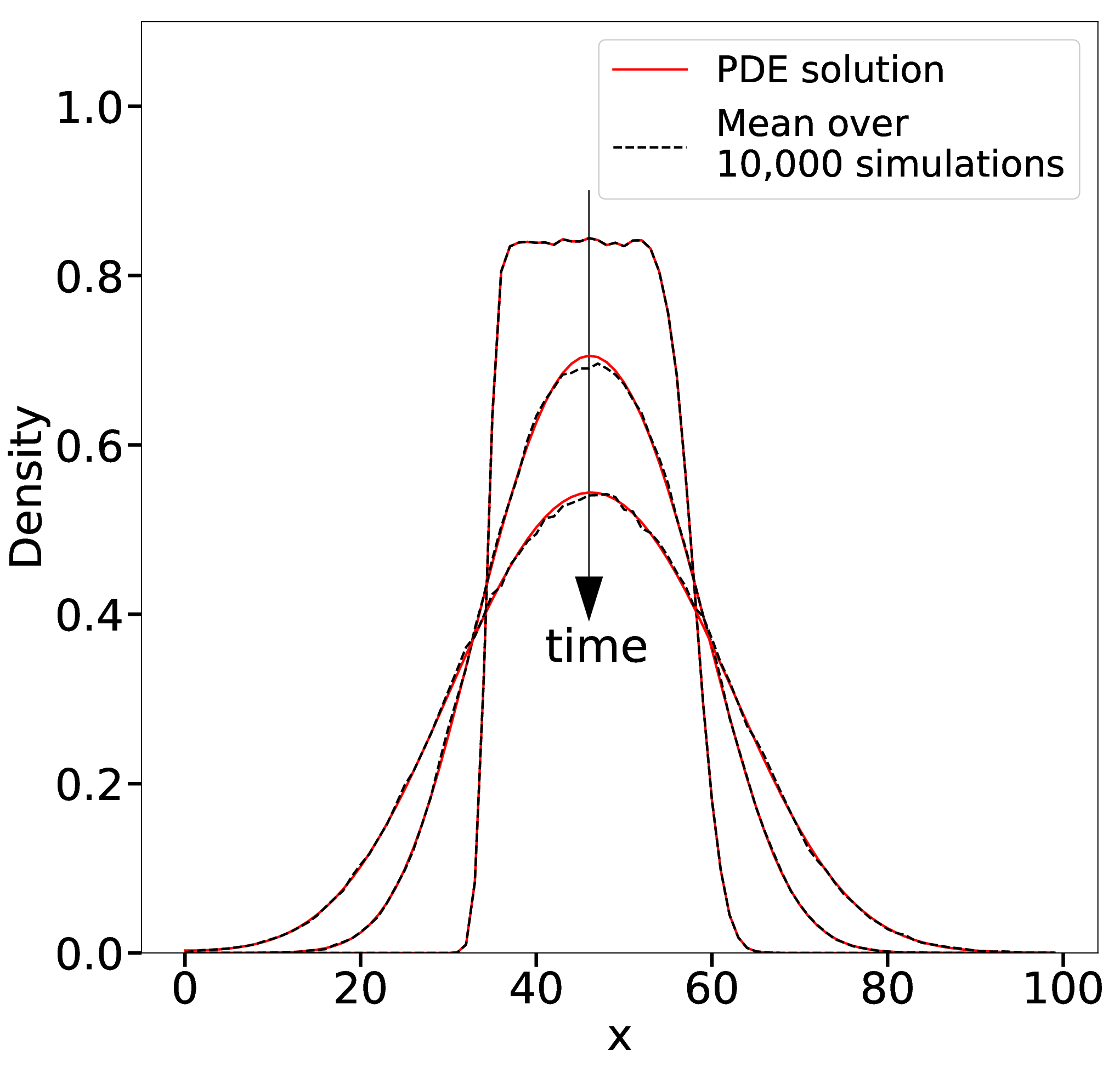}
\end{center}
\caption{Comparing the behaviour of the off-lattice ABM, averaged over 10,000 repeats with the numerical solution of the corresponding PDE \eqref{eq:PDE_off_lattice_d0}. We use parameters $N=20$, $R=0.17$, $p=25$ and $d=0.1$. For the initial conditions, in each repeat we place the leftmost agents at a position sampled from $N(35,1)$. We then place the remaining agents so that the distance between the centres of two adjacent agents is distributed uniformly on $[2R,12R]$. The PDE is solved using an explicit finite difference scheme, where the initial conditions are obtained by inputting the averaged ABM data at $t=0$. Density profiles are compared at $t=0$, $t=200$ and $t=500$.}
\label{fig:offlat_comparison}
\end{figure}

Let $P_r^i(x,t)$ be the probability that an agent is blocking a potential rightwards move by agent $i$ with centre at position $x$. In other words, the probability that an agent is present in the region $[x+2R,x+2R+d]$, so
\begin{equation}
P_r^i(x,t)=\sum_{j\neq i}\int_{2R}^{2R+d}C_j(x+s,t)\ ds.
\label{eq:p_r}
\end{equation}
Similarly, we can define the probability that an agent is present in the region $[x-2R-d,x-2R]$ as
\begin{equation}
P_l^i(x,t)=\sum_{j\neq i}\int_{-2R}^{-2R-d}C_j(x+s,t)\ ds.
\label{eq:p_l}
\end{equation}
We can now re-write the mean-occupancy equation \eqref{eq:Model_1_PME} more compactly as
\begin{align}
C_i(x,t+dt)&=C_i(x,t)+\frac{p dt}{2}C_i(x-d,t)\big(1-P_r^i(x-d,t)\big) \nonumber \\
&+\frac{p dt}{2}C_i(x+d,t)\big(1-P_l^i(x+d,t)\big) \nonumber \\
&-\frac{p dt}{2}C_i(x,t)\big(1-P_r^i(x,t)\big) \nonumber \\
&-\frac{p dt}{2}C_i(x,t)\big(1-P_l^i(x,t)\big).
\label{eq:PME_offlat}
\end{align}
Continuing to follow \citet{dyson2012mli} we take the appropriate limits in time and space to obtain the following PDE:
\begin{align}
\D{C_i}{t}&=\frac{p C_i}{2}\Big(P_r^i(x,t)+P_l^i(x,t)-P_r^i(x-d,t)-P_l^i(x+d,t)\Big) \nonumber \\
&+\frac{p d}{2}\D{C_i}{x}\Big(P_r^i(x-d,t)-P_l^i(x+d,t)\Big)+\frac{p d^2}{4}\DD{C_i}{x}\Big(2-P_r^i(x-d,t)-P_l^i(x+d,t)\Big)+\mathcal{O}(d^2).
\label{eq:PME_offlat_2}
\end{align}
Now we can also obtain approximations for the probabilities $P_r^i(x,t)$ and $P_l^i(x,t)$ by Taylor expanding $P_i(x-d,t)$ and $P_i(x+d,t)$ about $x$ to second-order and integrating the resulting polynomial.
\begin{align}
&P_r^i(x,t)\approx\sum_{j\neq i}\Big[dC_j+\frac{d}{2}(4R+d)\D{C_j}{x}+\frac{d}{6}(12R^2+6Rd+d^2)\DD{C_j}{x}\Big]  \label{eq:p_r_x}, \\
&P_r^i(x-d,t)\approx\sum_{j\neq i}\Big[dC_j+\frac{d}{2}(4R-d)\D{C_j}{x}+\frac{d}{6}(12R^2-6Rd+d^2)\DD{C_j}{x}\Big]  \label{eq:p_r_x-d}, \\
&P_l^i(x,t)\approx\sum_{j\neq i}\Big[dC_j-\frac{d}{2}(4R+d)\D{C_j}{x}+\frac{d}{6}(12R^2+6Rd+d^2)\DD{C_j}{x}\Big] \label{eq:p_l_x}, \\
&P_l^i(x+d,t)\approx\sum_{j\neq i}\Big[dC_j-\frac{d}{2}(4R-d)\D{C_j}{x}+\frac{d}{6}(12R^2-6Rd+d^2)\DD{C_j}{x}\Big] \label{eq:p_l_x+d} .
\end{align}
Substituting in the approximations \eqref{eq:p_r_x}-\eqref{eq:p_l_x+d} into equation \eqref{eq:PME_offlat_2} and neglecting terms which have combined order in $R$ and $d$ greater than $3$, we obtain the following PDE 
\begin{align*}
\D{C_i}{t}&=\frac{p d^2}{2}4RC_i\sum_{j\neq i}\DD{C_j}{x}+\frac{p d^2}{2}\big(4R-d\big)\D{C_i}{x}\sum_{j\neq i}\D{C_j}{x}+\frac{p d^2}{2}\DD{C_i}{x}-\frac{p d^3}{2}\DD{C_i}{x}\sum_{j\neq i}C_j.
\end{align*}
We assume that all the agents were initialised with position chosen from the same probability distribution. Since all agents move according to identical rules, this means we can replace $C_i(x,t)$ with $C_j(x,t)$ for any $i$ and $j$.
\begin{equation}
\D{C_i}{t}=\frac{p d^2}{2}\left(\DD{C_i}{x}+(N-1)(4R-d)\D{}{x}\Big(C_i\D{C_i}{x}\Big)\right).
\label{eq:PDE_off_lattice_single}
\end{equation}
This PDE describes the evolution of the density function of a single agent. We can use equation \eqref{eq:PDE_off_lattice_single} to derive a PDE for the total density of cells $C(x,t)=\sum_{i=1}^{N}C_i(x,t)=NC_i(x,t)$.
\begin{equation}
\D{C}{t}=\frac{p d^2}{2}\D{}{x}\Big[\Big(1+(4R-d)\frac{N-1}{N}C\Big)\D{C}{x}\Big].
\label{eq:PDE_off_lattice}
\end{equation}
Taking the limit as $d\rightarrow 0$ and $p\rightarrow \infty$ while keeping $p d^2$ constant gives 
\begin{equation}
\D{C}{t}=\hat{p}\D{}{x}\Big[\Big(1+4RC\frac{N-1}{N}\Big)\D{C}{x}\Big],
\label{eq:PDE_off_lattice_d0}
\end{equation}
where 
$$\hat{p}=\lim_{\substack{d\to 0 \\ p\to\infty}} p d^2.$$
In contrast to the on-lattice equivalent of this model (i.e. equation \eqref{eq:PDEpull} with $w=0$), the volume exclusion effects now enhance the diffusion coefficient via the density-dependent term $4RC(N-1)/N$.
When comparing the discrete ABM and the continuous PDE, rather than representing each agent as a hat function, we represent each agent by a Gaussian probability density function, with standard deviation equal to the radius of the agent (as in \citet{yates2015ipe}). This has the effect of smoothing the data so that we need run fewer repeats. Figure \ref{fig:offlat_comparison} demonstrates the fit between the averaged ABM and the PDE is very good.

\subsection{Contact forming model} \label{sec:Model_2}
\begin{figure}[h]
\begin{center} 
\subfigure[] {
\includegraphics[width=.45\textwidth]{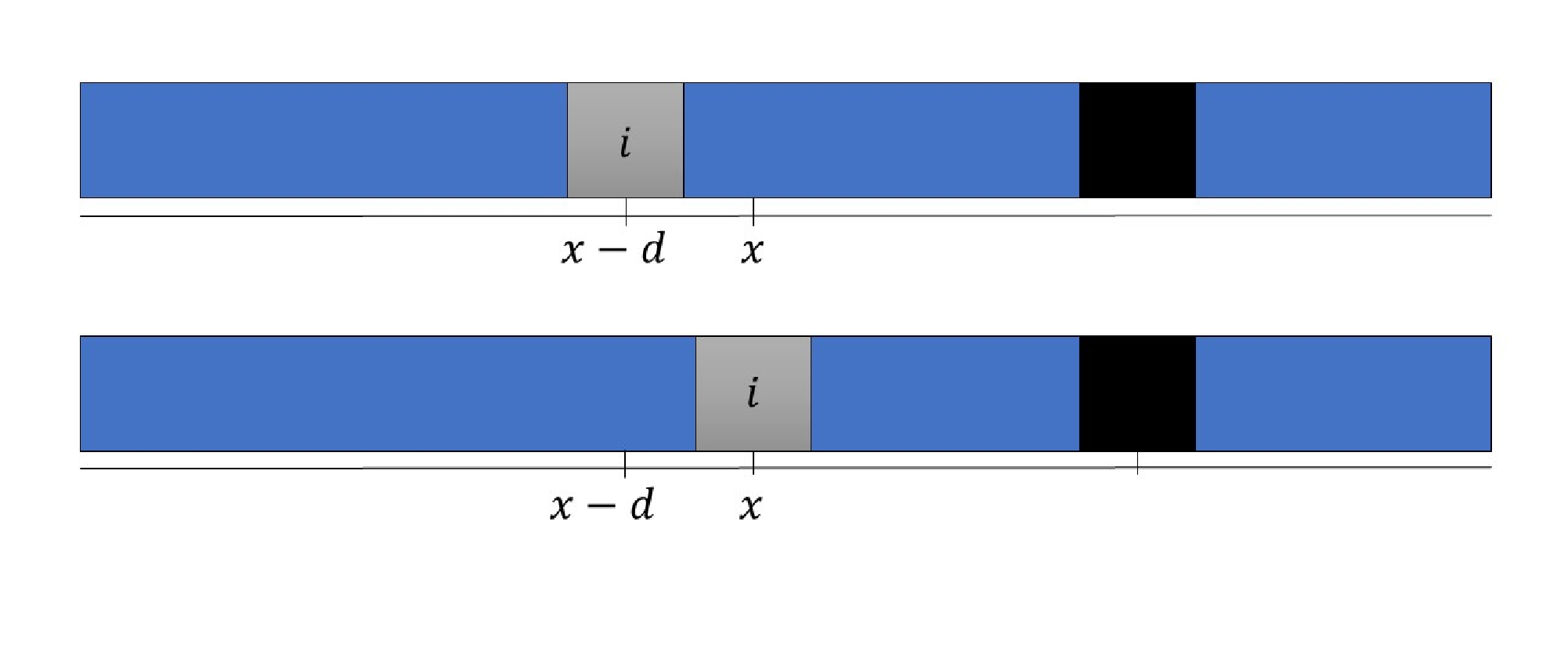}
\label{fig:off_lattice_full}
}
\subfigure[]{
\includegraphics[width=.45\textwidth]{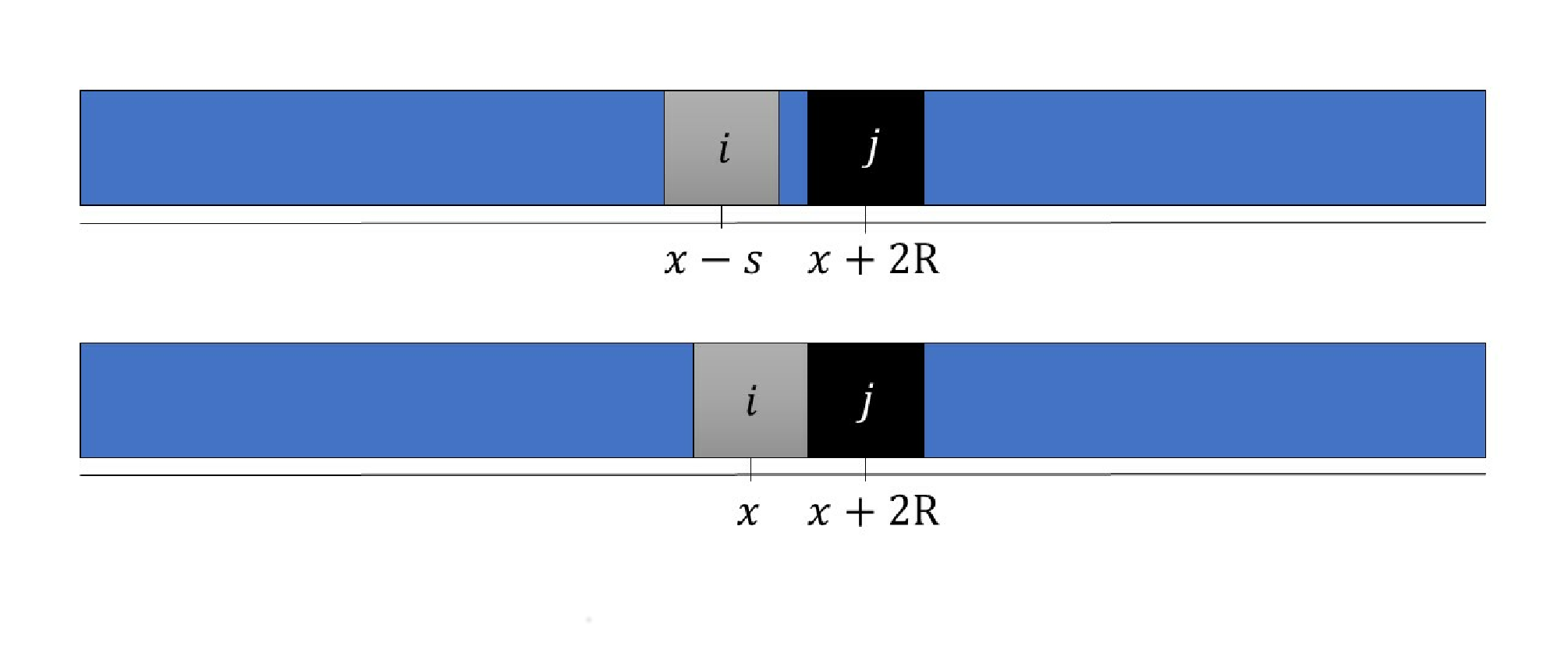}
\label{fig:off_lattice_rest}
}
\end{center}
\caption{The two ways in which agent $i$ may move into position $x$. Panel \subref{fig:basic_ex_schematic} shows agent $i$ originally centred at position $x-d$ (upper) moving distance $d$ into position $x$ in the absence of any blocking agent (lower). Panel \subref{fig:pulling_schematic} shows agent $i$ originally centred at $x-s$ (upper, where $s<d$) moving a restricted distance $s$ so that it ends up in contact with a blocking agent $j$ centred at $x+2R$ (lower). }
\label{fig:offlat_cell_schematics}
\end{figure}

We now adapt the model so that, if a proposed move would result in an overlap, the agent moves so that it is in contact with the blocking agent. The mean-occupancy equation for this model then reads as follows:
\begin{align}
C_i(x,t+dt)-C_i(x,t)&=-\frac{p dt}{2}C_i(x,t)\big(1-\sum_{j\neq i}C_j(x-2R,t)\big) \nonumber \\
&-\frac{p dt}{2}C_i(x,t)\big(1-\sum_{j\neq i}C_j(x+2R,t)\big) \nonumber \\
&+\frac{p dt}{2}C_i(x-d,t)\Big(1-\sum_{j\neq i}\int_{2R}^{2R+d}C_j(x-d+s,t)ds\Big)  \nonumber\\
&+\frac{p dt}{2}C_i(x+d,t)\Big(1-\sum_{j\neq i}\int_{-2R-d}^{-2R}C_j(x+d+s,t)ds\Big)\nonumber \\
&+\frac{p dt}{2}\sum_{j\neq i}C_j(x+2R,t)\int_{0}^{d}C_i(x-s,t)ds\nonumber \\
&+\frac{p dt}{2}\sum_{j\neq i}C_j(x-2R,t)\int_{0}^{d}C_i(x+s,t)ds \nonumber\\
&-\frac{p dt}{2}C_i(x,t)\sum_{j\neq i}C_j(x-2R,t)\nonumber \\
&-\frac{p dt}{2}C_i(x,t)\sum_{j\neq i}C_j(x+2R,t).
\label{eq:PME_offlat_cont} 
\end{align}
The first term represents losing occupancy at position $x$ if agent $i$, initially centred at position $x$ attempts to move rightwards and there is no agent in contact with agent $i$ on its right side. Agent $i$ is then free to move rightwards, even if only by a small amount, and does so with probability $p dt/2$. This move is depicted in Figure  \ref{fig:offlat_cell_schematics} \subref{fig:off_lattice_rest}. The second term represents the equivalent for a leftwards move, i.e. agent $i$ is able to move leftwards out of position $x$.

The third term represents the gain in occupancy at position $x$ if agent $i$, initially centred at $x-d$, attempts to move rightwards and there is no part of another agent in the region $[x-d+R,x+R]$. Therefore agent $i$ is free to jump a distance $d$ to the right into position $x$, with probability $p dt/2$. This situation is visualised in Figure  \ref{fig:offlat_cell_schematics} \subref{fig:off_lattice_full}. The fourth term is the mirror image of the third term: agent $i$ moves leftwards from position $x+d$ to $x$. The fifth term captures the fact that we can also gain occupancy at position $x$ if agent $i$ has its centre in the region $[x-d,x]$ and there is another agent whose centre is at exactly $x+2R$ blocking agent $i$'s rightwards move. This means agent $i$ moves a restricted distance (less than $d$) and ends up in contact with the agent whose centre is at $x+2R$. The sixth term is then the mirror image of this for leftwards moves. The final two terms are to prevent double counting. If we are in a situation where agent $i$ is at position $x$ and there is another agent in contact with it at position $x+2R$ the fifth term would cause us to erroneously gain occupancy at position $x$ despite the fact that no change in occupancy could occur as the agent cannot move to the right. In order to avoid this we subtract the density evaluated at the lower limit of the integrals. 

Following the procedure outlined above (see Section \ref{SUPP-sec:Model_2_PDE} supplementary material), the PDE for the average agent density, $C=\sum_{i=1}^NC_i=NC_i$, is given by
\begin{equation}
\D{C}{t}=\hat{p}\D{}{x}\Big[\Big(1+2RC\frac{N-1}{N}\Big)\D{C}{x}\Big],
\label{eq:PDE_offlat_cont_d0}
\end{equation}
where $\hat{p}=\lim_{d\to 0,p\to\infty} p d^2$ is the diffusive limit, as in equation \eqref{eq:PDE_off_lattice_d0} (see section \ref{SUPP-sec:Model_2_PDE} of the supplementary material for the derivation). Comparing equation \eqref{eq:PDE_offlat_cont_d0} with equation \eqref{eq:PDE_off_lattice_d0} suggests that, according to the PDEs, the effect of allowing agents to form contacts rather than aborting overlapping moves is to slightly reduce the effective diffusion coefficient. Intuitively, allowing agents to form contacts has the effect of bringing agents together which were not previously touching, effectively reducing the diffusive spread of agents.

\subsection{Pulling at a distance} \label{subsec:offlat_pulling_distance}
In order to incorporate pulling we assume that agents are able to pull neighbours at a distance. This is the off-lattice analogue of the on-lattice model described is section \ref{subsec:pulling_distance}. We adapt the aborting moves model in section \ref{sec:Model_1} as follows. We introduce a pulling distance, $l$, such that if agent $i$ is chosen to move right and there is another agent $j$ whose right edge is within $l$ of agent $i$'s left edge then, with probability $Q$, both agents move a distance $d$ to the right. The mean-occupancy equation is included as equation \eqref{SUPP-eq:offlat_pulling_distance_PME} in the supplementary material. 

Taylor expanding appropriate terms, integrating, re-arranging and discarding higher-order terms in the mean-occupancy equation gives the, somewhat complex, PDE for the probability distribution of agent $i$
\begin{align*}
\D{C_i}{t}&=-\frac{p}{2}\left[Q(4R+l)(N-1)\left(C_i\D{C_i}{x}+\left(\DD{C_i}{x}\right)^2\right)\right]dl \\
&+\frac{p}{2}\left[\DD{C_i}{x}+(4R+2lQ)(N-1)\left(C_i\DD{C_i}{x}+\left(\D{C_i}{x}\right)^2\right)\right]d^2 \\
&-\frac{p}{2}\left[(N-1)\left(C_i\DD{C_i}{x}+\left(\D{C_i}{x}\right)^2\right)\right]d^3 \\
&=\frac{p d^2}{2}\DD{C_i}{x}+d(N-1)\D{}{x}\left(C_i\D{C_i}{x}\right)\left(-(4R+l)Ql+(4R+2lQ)d-d^2\right).
\end{align*}
In order to take the diffusive limit, we assume that the pulling distance is proportional to the movement distance, so $l=dk$ for some $k>0$. The PDE for the average occupancy of agent $i$ then simplifies to
\begin{align*}
\D{C_i}{t}=\frac{p d^2}{2}\D{}{x}\left[1+\left(N-1\right)\left(4R\left(1-Qk\right)-d\left(1+Qk^2-2Qk\right)C_i\right)\D{C_i}{x}\right].
\end{align*}
For the total agent density  $C=\sum_{i=1}^NC_i$, we obtain
\begin{equation}
\D{C}{t}=\frac{p d^2}{2}\D{}{x}\left[\Big(1+\frac{N-1}{N}\left(4R\left(1-Qk\right)-d\left(1+Qk^2-2Qk\right)\right)C\Big)\D{C}{x}\right].
\label{eq:PDE_offlat_pull_d}
\end{equation}
Taking the diffusive limit as in equation \eqref{eq:PDE_off_lattice_d0} gives
\begin{equation}
\D{C}{t}=\hat{p}\D{}{x}\left[\Big(1+\frac{N-1}{N}4RC\left(1-Qk\right)\Big)\D{C}{x}\right],
\label{eq:PDE_offlat_pull_d0}
\end{equation}
where $\hat{p}$ is defined as before.

We note that this PDE predicts that the effect of pulling is to decrease the effective diffusion coefficient, causing agents to disperse more slowly. This prediction differs from the corresponding on-lattice model described by equation \eqref{eq:PDEpull_d}. In the on-lattice case, pulling at a distance enhances the diffusion coefficient, except at very low densities.

\subsection{Error comparison}\label{subsection:off_lattice_error_comparison}
\begin{figure}[h!!!]
\begin{center} 
\subfigure[]{
\includegraphics[width=0.31\textwidth]{./offlat_comp_plot.eps}
\label{subfigure:offlat_ex_pde_sim}
}
\subfigure[]{
\includegraphics[width=0.31\textwidth]{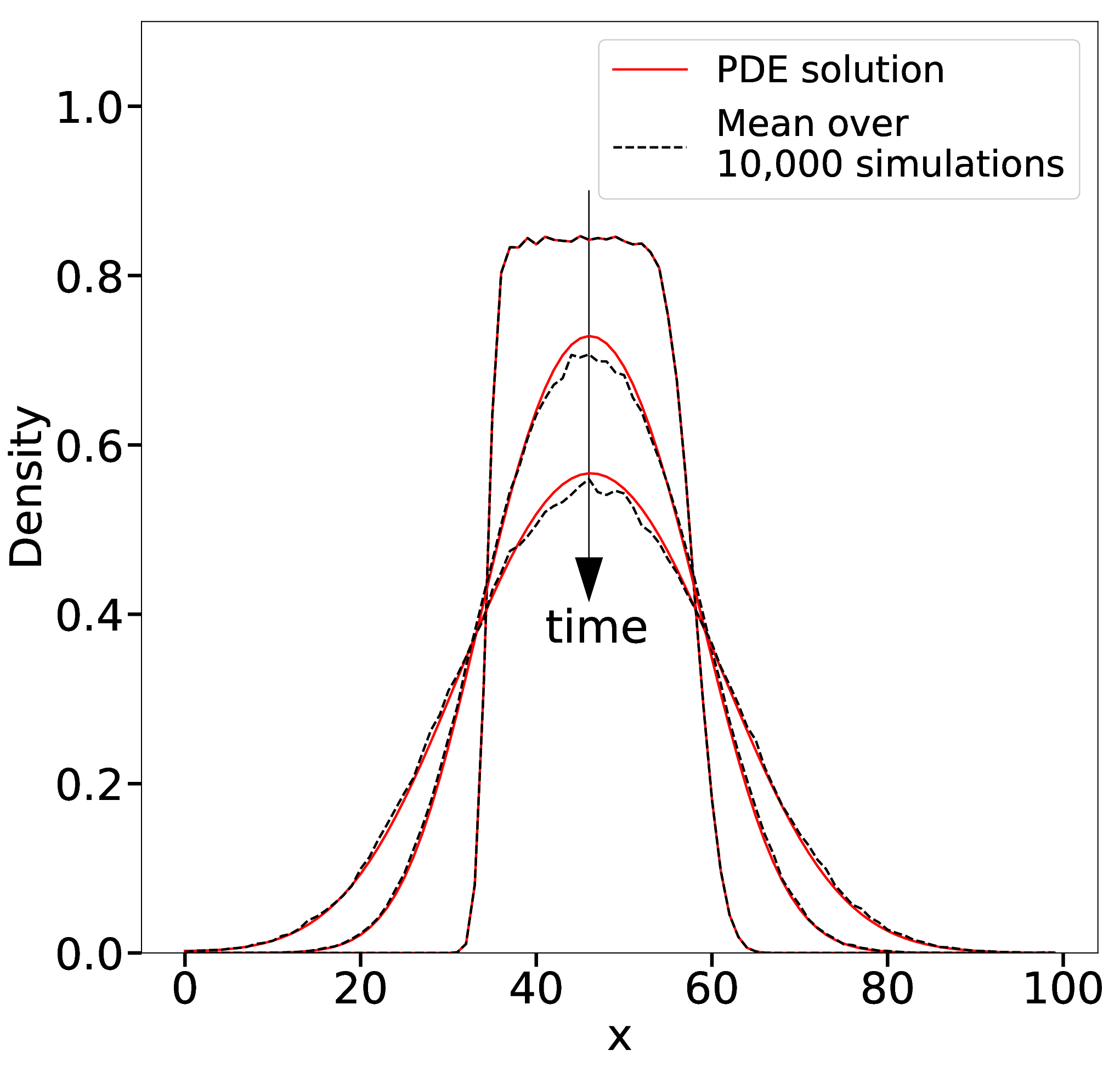}
\label{subfigure:offlat_con_pde_sim}
}
\subfigure[]{
\includegraphics[width=0.31\textwidth]{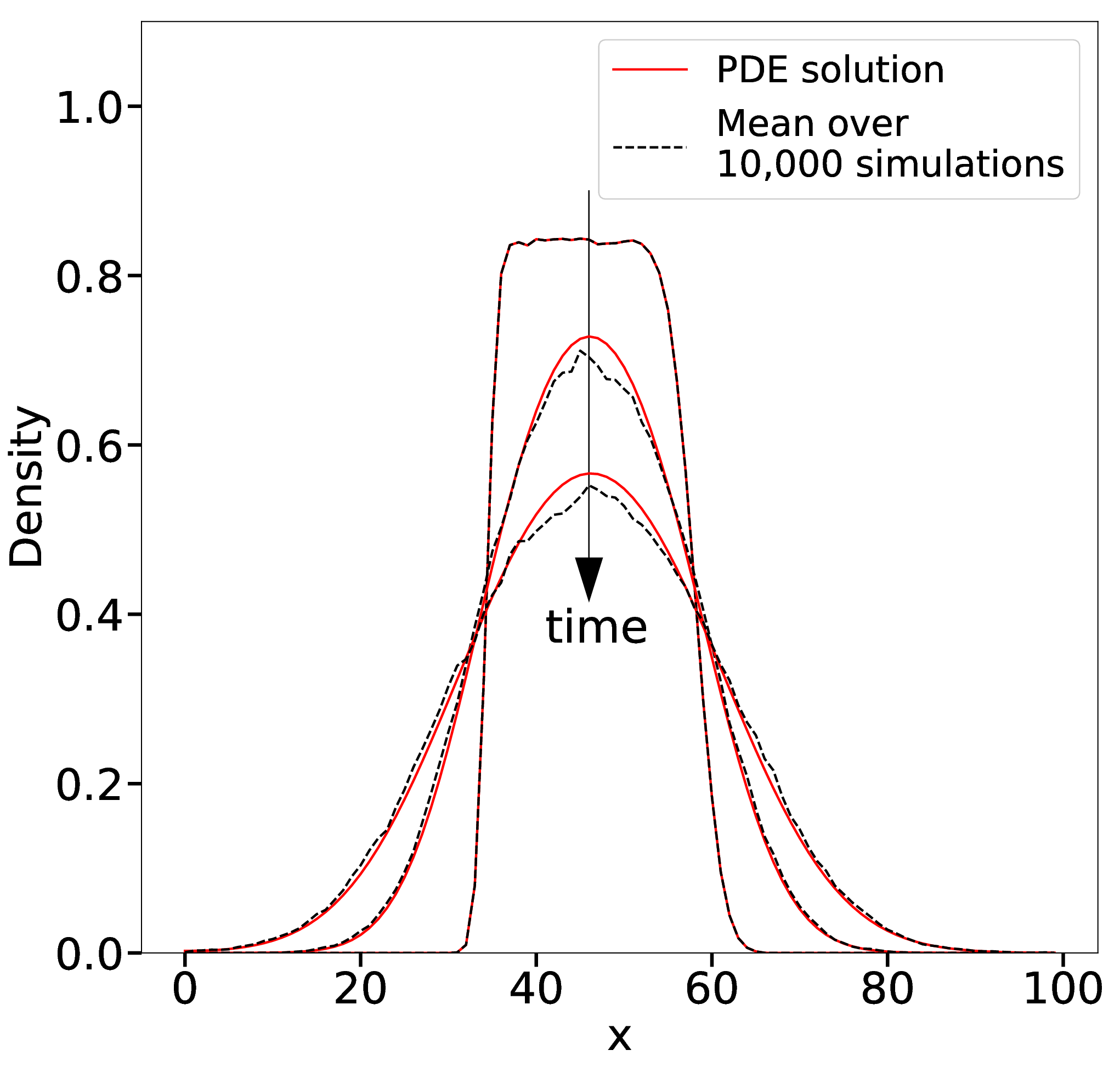}
\label{subfigure:offlat_pulld_pde_sim}
}
\end{center}
\caption{Comparing the behaviour of the off-lattice ABMs, averaged over 10,000 repeats with the numerical solution of the corresponding PDE. For all three models we use parameters $N=20$, $R=0.17$, $p=25$ and $d=0.1$. Simulations are carried out on the domain $[0,100]$. Initial conditions are handled in the same way as in Figure \ref{fig:offlat_comparison}. Panel \subref{subfigure:offlat_ex_pde_sim} represents simulated results for the model in which overlapping moves are aborted. Panel \subref{subfigure:offlat_con_pde_sim} represents simulated results for the model in which we shorten overlapping moves so agent form contacts. Panel \subref{subfigure:offlat_pulld_pde_sim} represents results for the model in which we abort overlapping moves and allow pulling at a distance, with parameter $Q=1$, $k=1/2$. The PDEs are solved using an explicit finite difference scheme. Density profiles are compared at $t=0$, $t=200$ and $t=500$.}
\label{fig:offlat_sim_vs_pde}
\end{figure}

In this section we compare the averaged behaviour of the off-lattice ABMs with the solution of their corresponding PDEs. The comparisons are shown in Figure \ref{fig:offlat_sim_vs_pde}. The first model, in which overlapping moves are aborted, is represented very well by its mean-field PDE \eqref{eq:PDE_off_lattice}. For the second model, in which agents move to be in contact with one another, and the third model in which agents can pull each other at a distance) there is a slightly larger discrepancy between the averaged ABMs and their corresponding PDEs given by equations \eqref{eq:PDE_offlat_cont_d0} and \eqref{eq:PDE_offlat_pull_d}, respectively. 

\begin{figure}[h!]
\begin{center} 
\includegraphics[width=0.7\textwidth]{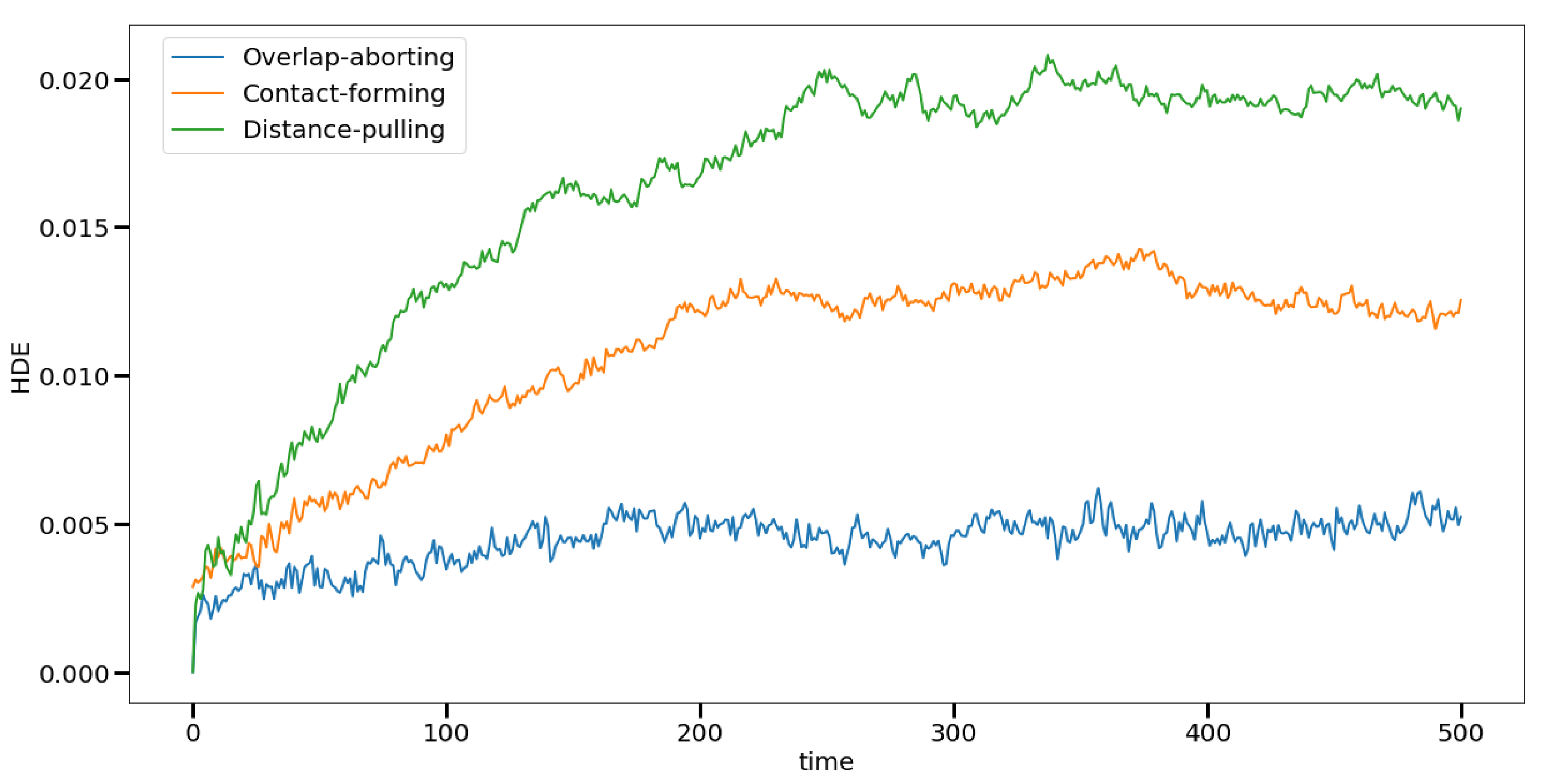}
\end{center}
\caption{Comparison of the evolution of the HDE comparing each model to its respective PDE. We use the same domain and parameters as for Figure \ref{fig:offlat_sim_vs_pde}.  We observe that the more complex the model, the larger the discrepancy between the ABM and the PDE.}
\label{fig:offlat_HDE}
\end{figure}

The agreement between the ABMs and the PDEs is further quantified in Figure \ref{fig:offlat_HDE} in which we consider the evolution of the HDE between the computed PDE solution and the smoothed averaged density profiles of the ABM evaluated at the grid points of the PDE solution.

As in the on-lattice case, one explanation for the discrepancies is the assumption that the occupancy of nearby positions in the domain are independent. This assumption becomes less valid the more that agents interact with each other so it is reasonable that the discrepancy between the PDE and the ABM is larger for the models where more interaction takes place. There are other possible sources discrepancy which could also be contributing. For example, we used a second-order Taylor expansion making use of the assumption that higher-order terms disappear in the limit as $d\rightarrow0$. The validity of this assumption relies on all spatial derivatives being small, which might not be the case. 

\section{Discussion}\label{chapter:discussion}

In this work we have addressed the lack of either ABM or continuum models which successfully incorporate cell-cell pulling. We began by considering simple on-lattice models and steadily increased the complexity of the cell-cell interactions in order to represent several different possible pulling mechanisms. For each of these variants we derived a corresponding mean-field PDE. In all cases, the corresponding PDE was a non-linear diffusion equation with a density-dependent diffusion coefficient. In almost all cases, the effect of pulling was to augment the diffusion coefficient so that cells disperse more quickly, with different pulling mechanisms augmenting the diffusion coefficient to differing degrees. These comparisons demonstrated that incorporation of cell-cell pulling can have a significant impact on the behaviour of cells in both agent-based and population-level models of the same phenomenon. 

In the introduction we noted a number of other studies which have derived non-linear diffusion equations from microscopic interactions between cells moving on a lattice. In Section \ref{subsec:Comparison_with_pushing} we explicitly compared the effects of cell-cell pulling with cell-cell pushing. It would be interesting, in future work, to perform a more systematic comparison between the other models incorporating cell-cell interactions which exist in the literature. One important feature of the diffusion coefficients we derived is that they cannot become negative for any parameter values, meaning the PDE remains well-posed. This is in contrast to the diffusion coefficient derived for cell-cell adhesion in \cite{thompson2012mcm} and some of the interactions considered in \cite{fernando2010nde}. A common finding between this work and many of the others is that the correspondence between the ABM and the mean-field PDE becomes poorer for stronger or more complex interactions. 

Beyond the literature on modelling collective cell migration, non-linear diffusion equations have also been used by mathematical ecologists to model the dispersal of species \citep{gurney1975rip, mendez2012ddp}. In these cases, the form of the density-dependent diffusion coefficient is chosen phenomenologically rather than derived from agent-agent interactions. This work may help to shed light on the types of interactions between species which give rise to such non-linear diffusion models. 

By considering off-lattice models of cell migration we investigated how the type of ABM chosen impacts upon the population-level behaviour. We found two important differences between the on-lattice and off-lattice cases. Firstly, for the pulling mechanisms we considered, the population-level effects of pulling in the ABM are almost non-existent in the off-lattice case. This is in contrast to the on-lattice case for which cell-cell pulling causes agent to disperse more quickly. We also found that the agreement between the off-lattice ABM and the corresponding PDE breaks down quite quickly for models with non-trivial agent to agent interaction. 

The differences between the range of pulling models we have presented here emphasise the importance of faithfully representing the underlying biological process. We have seen that different types of cells employ different types of pulling mechanism, and some types of cells are better suited to the lattice approximation than others. All of this needs to taken into account when choosing a model. \citet{osborne2017cib} provide an investigation of model selection in the context of computational modelling collective cell migration. 

 Another practical question which arises when applying such models is how they can be parameterised. It may be possible to approximate parameter values by direct observation of migrating cells. Another approach is to use a probabilistic methodology such as Approximate Bayesian Computation \citep{ross2017uab,johnston2014isa} for which posterior distributions of parameter values are estimated by comparing large numbers of forwards model simulations with data.

This initial exploratory work provides a platform from which further investigations into the effects of cell-cell pulling can be launched. For example, we studied the simplest possible case of cells which are able to pull at a distance. There may be more complex behaviour in models involving pulling over larger distances and of multiple agents. Another possibility is to consider multiple species of cells, some of which are capable of pulling whilst others are not. 

We found that the correspondence between the ABMs and the mean-field PDEs becomes poorer with increased cell-cell interaction complexity. The assumption that the occupancy of neighbouring sites is uncorrelated becomes increasingly invalid as the complexity of the interactions increases. Another possible area to which this work might be extended is the use of spatial correlation functions in order to derive more accurate PDEs for on-lattice models \cite{baker2010cmf,markham2013smi}.

The method we have used to derive a continuum PDE from an off-lattice ABM is by no means the only possible method. Other methods exist and have been used successfully to derive continuum analogues which provide accurate representations of the mean-field behaviour of volume-excluding AMBs \citet{bruna2012eve,taylor2014mmt,franz2016hsi,plank2013lfm,plank2012mcc}. It is possible that some of these methods might provide more accurate continuum models than our simple mean-field representations of cell-cell pulling.
 


\end{document}